\documentclass{aa}
\usepackage[colorlinks=true,citecolor=blue,anchorcolor=blue,filecolor=blue,linkcolor=blue]{hyperref}
\usepackage{comment}
\usepackage{soul} 
\usepackage{graphicx}
\usepackage{amssymb,amsmath}
\usepackage{latexsym}
\usepackage{mathtools}
\usepackage{epsfig}
\usepackage{natbib}
\usepackage{dsfont}
\usepackage{xcolor}
\usepackage{txfonts}
\usepackage{mathrsfs}
\usepackage[super]{nth}
\usepackage{multirow}
\usepackage{floatrow}
\usepackage{arydshln}

\DeclareMathOperator{\sech}{sech}
\begin{document}

   \title{Galactic tides in the Solar System within a non-axisymmetric\\Milky Way model adjusted to Gaia data}
%
   \author{Alexandre Bougakov$^1$
          \and 
          Yassin Rany Khalil$^{2,3}$
          \and
          Marc Fouchard$^1$
          \and
          Melaine Saillenfest$^1$
          \and
          Benoit Famaey$^3$
          \and
          Paola Di Matteo$^4$
          \and
          Misha Haywood$^4$
          }
   \authorrunning{Bougakov et al.}
   \institute{$^1$LTE, Observatoire de Paris, Université PSL, Sorbonne Université, Université de Lille, LNE, CNRS, 61 Avenue de l'Observatoire, F-75014 Paris, France\\$^2$Università degli Studi di Firenze, Dipartimento di Fisica e Astronomia, Via G. Sansone 1, I-50019, Sesto Fiorentino, Italy\\$^3$Université de Strasbourg, CNRS UMR 7550, Observatoire astronomique de Strasbourg, F-67000 Strasbourg, France\\$^4$LIRA, Observatoire de Paris, Université PSL, Sorbonne Université, Université Paris Cité, CY Cergy Paris Université, CNRS, F-92190 Meudon, France\\ \email{alexandre.bougakov@obspm.fr}}
   \date{Received 2026-03-25 / Accepted 2026-06-26}


  \abstract
  {Galactic tides are the external differential forces felt by extended systems immersed in the Galactic potential. They are known to play a determinant role in the dynamics of comets in the Oort cloud.}
  {We aim to establish the strength of the Galactic tides driven by the non-axisymmetric potential of the Milky Way revealed by the Gaia mission, and how this strength evolves along the Galactic trajectory of the Solar System.}
  {We derived expressions for the Galactic tide parameters independently of any simplified trajectory for a star in the Galaxy or any specific symmetry for the Galactic potential. We obtained a set of six parameters, $G_1$ to $G_6$, that quantify the influence of tides, rather than the traditional three parameters $G_1$ to $G_3$. Using the most up-to-date model for the Galactic potential, extended here to three dimensions, we studied the time evolution of these parameters along the trajectory of the Sun using a statistical approach.}
  {Even for Galactic trajectories featuring modest radial and vertical excursions, such as those investigated here, the dominant Galactic parameter, $G_3$, is found to vary by an order of magnitude along the solar trajectory. The parameter $G_1$ features even more significant variations over time produced by the Galactic spiral arms. In general, $G_1$ and $G_2$ reach up to about one half and one quarter, respectively, of the value of $G_3$ along the solar trajectory, whilst parameters $G_4$ to $G_6$ reach up to one tenth the value of $G_3$. We provide a tabulated time evolution of all parameters, for use in Solar System studies.}
  {Contrary to what is often assumed, all Galactic tide parameters vary widely along the trajectory of the Solar System within the Galaxy. One can expect two consequences of such variations: First, $G_3$ directly affects the flux of observable long-period comets and the extent of the fossilised Sednoids region; second, $G_1$ and $G_2$, and to a lesser extent, $G_4$ to $G_6$, break the integrability of the dynamics, with a possible impact on the long-term structure of the Oort cloud. Additionally, the new parameters $G_4$ to $G_6$, while small for the Solar System, may have a strong impact on extrasolar systems with a large out-of-the-plane excursion in the Galaxy.}

  \keywords{}

  \maketitle

\section{Introduction}\label{sec:intro}
The Oort cloud is a distant spherical reservoir of bodies located at the periphery of the Solar System~\citep{Oort_1950}. It is widely considered to be the primary source of long-period comets~\citep[see e.g.][]{Kaib-Volk_2024}, with orbital periods exceeding~$200$~yr. These comets are believed to be primordial relics of the solar nebula and thus offer an exceptional opportunity to investigate the formation and evolutionary history of the Solar System~\citep[see e.g.][]{Morbidelli_2005,Fouchard-Emelyanenko-Higuchi_2020}. The most distant trans-Neptunian objects (TNOs) are expected to preserve a record of the Solar System's formation and evolution over more than four billion years, including the Sun's birth cluster, planetary migration, stellar encounters, the evolution of the Sun's Galactic environment, and so on~\citep[see][]{Dones-Weissman-Levison-Duncan_2004a,Brasser-Morbidelli_2013,Fouchard-Higuchi-Ito-Maquet_2018,Fouchard-Emelyanenko-Higuchi_2020,Morbidelli-Nesvorny_2020,Fouchard-Higuchi-Ito_2023}. Furthermore, some TNOs have orbits that suggest the existence of an unknown planet in the far reaches of the Solar System~\citep[see e.g.][]{Trujillo-Sheppard_2014,Batygin-Brown_2016,Batygin-Adams-Brown-Becker_2019,Sheppard-Trujillo-Tholen-Kaib_2019}. 

The mechanisms responsible for the injection and ejection of comets are governed by several gravitational processes, including the Galactic tides --- first invoked by~\cite{Chebotarev_1966}, then rediscovered by~\cite{Byl_1983}, and later refined by~\cite{Heisler-Tremaine_1986} --- as well as by sporadic encounters with massive objects such as stars~\citep[see e.g.][]{Rickman-Fouchard-Froeschle-Valsecchi_2008,Fouchard-Rickman-Froeschle-Valsecchi_2017} and molecular clouds~\citep[see e.g.][]{Clube-Napier_1986,Torbett_1986b,Heisler_1990}. These two kinds of external forces have very distinct effects on the orbits of comets, and therefore are modelled separately: Galactic tides produce smooth quasi-integrable orbital dynamics, whereas encounters produce stochastic impulses. In this paper, we only focus on Galactic tides. For a comprehensive overview of the different mechanisms affecting the orbits of TNOs, see~\cite{Saillenfest_2020}.

\cite{Heisler-Tremaine_1986} made a major contribution to the study of the long-term dynamics of TNOs influenced by Galactic tides, thereby establishing the foundations of the field. As a result, subsequent studies have naturally built upon this work and made extensive use of its results (see most references in this introduction). The work of these authors on quantifying the impact of Galactic tides on the Solar System has also had consequences for related areas: the exoplanet orbits~\citep[][]{Veras-Evans_2013}, the dynamics of wide binary stars with exoplanets~\citep[][]{CorreaOtto-GilHutton_2017}, the dynamics of wide stellar triples~\citep[][]{Grishin_Perets-2022}, the formation of hot Jupiters in wide binaries~\citep[][]{Grishin-Winter-Alvarado-Montes_2025}. However, the work of~\cite{Heisler-Tremaine_1986} was developed in a context that differs substantially from the current understanding of the Sun's Galactic environment and from the quality of observational data available in recent decades. In fact, the canonical model of Galactic tides built by~\cite{Heisler-Tremaine_1986} relies on important simplifications: The trajectory of the Sun is assumed to be circular in the Galactic plane, and the Galactic potential is assumed to be independent of time and have axial and north-south symmetries. These simplifications were primarily intended for order-of-magnitude estimates. In contrast, recent data from the Gaia mission \citep{GaiaCollaboration_2018, GaiaCollaboration_2023} have revealed the complexity of the phase-space structure of the Galaxy in exquisite detail \citep[see e.g.][for a review]{HuntVasiliev} and its connection to the non-axisymmetries of the Galactic disc, namely the bar and spiral arms \citep[e.g.,][]{Monari2019, Khoperskov-Gerhard-DiMatteo-Haywood-Katz-Khrapov-Khoperskov-Arnaboldi_2020,Laporte2020,Clarke,Hunter2024,Dillamore2025,Kalda2025,Khoperskov2025,Hamilton2026}, meaning that the Sun's motion within the Galaxy and the variation of its gravitational field and density of its environment are actually much more complex than in a simplified axisymmetric model. Recently, \cite{Khalil_2025} provided a realistic two-dimensional non-axisymmetric model of the Milky Way disc, adjusted to the observed local velocity distribution around the Sun as well as to the map of median galactocentric stellar radial velocities as a function of position in the Galactic plane, as measured with Gaia. This Galactic model, once extended to three dimensions, therefore will open the way for numerical integrations of the Sun's possible trajectories within the Galaxy and for estimating the related Galactic tide parameters acting on the Solar System.

\cite{Heisler-Tremaine_1986} and~\cite{Delsemme_1987} have shown that the flux of observable comets coming from the Oort cloud strongly depends on the normal component of these Galactic tides (i.e. perpendicular to the Galactic plane) acting on the Solar System. As investigated by~\cite{Gardner-Nurm-Flynn-Mikkola_2011}, the variations of these tides as the Sun moves in the Galaxy induce a correlated variation in the cometary flux. In addition, \cite{MartinezBarbosa-Jilkov-PortegiesZwart-Brown_2017} demonstrated that the number of stellar encounters depends on the type of trajectory the Sun followed from its birth to the present. According to the literature, the Sun's birth radius is very uncertain and estimated to lie between about 4.5 and 11~kpc~\citep[see e.g.][]{MartinezBarbosa-Brown-PortegiesZwart_2015,Frankel-Rix-Ting-Ness-Hogg_2018,Minchev-Anders-RecioBlanco-Chiappini-deLaverny-Queiroz-Steinmetz-Adibekyan-Carrillo-Cescutti-Guiglion-Hayden-deJong-Kordopatis-Majewski-Martig-Santiago_2018,Haywood-Snaith-Lehnert-DiMatteo-Khoperskov_2019,Baba-Tsujimoto-Saitoh_2024,Lu-Minchev-Buck-Khoperskov-Steinmetz-Libeskind-Cescutti-Freeman-Ratcliffe_2024,Dantas-Smiljanic-deSouza-Tissera-Magrini_2025}. If significant radial migration \citep[see][for dynamical mechanisms]{SB02,MF10} occurred along the Sun's lifetime, it could have had a major impact on TNO dynamics~\citep[see e.g.][]{Kaib-Roskar-Quinn_2011,Kaib-Volk_2024}. However, while radial migration must be present to some level in the Galactic disc, its magnitude, especially in the case of the Sun itself, is debated~\citep[see e.g.][]{Haywood-DiMatteo-Lehnert-Katz-Gomez_2013,Haywood-Snaith-Lehnert-DiMatteo-Khoperskov_2019,Halle-DiMatteo-Haywood-Combes_2015}. In any case, this sharply contrasts with studies that assume that the Sun moves on a circular orbit in the Galactic plane of a fully axisymmetric Milky Way~\citep[see e.g.][]{Heisler-Tremaine_1986,Wiegert-Tremaine_1999,Levison-Dones-Duncan_2001,Fouchard_2004,Brasser-Duncan-Levison_2008,Brasser-Higuchi-Kaib_2010,Breiter-Fouchard-Ratajczak_2008,Wang-Brasser_2014,Silsbee-Tremaine_2016,Namouni-Morais_2018,Saillenfest-Fouchard-Ito-Higuchi_2019,Clement-Kaib_2020,Fouchard-Higuchi-Ito_2023,Nesvorny-Bernardinelli-Vokrouhlicky-Batygin_2023,deSousa-Izidoro-Morbidelli-Nesvorny-Winter_2025}, because these assumptions mathematically imply no temporal variation in the Galactic tides and no temporal variation in the local mass density of the Galaxy in the solar neighbourhood, which limits the refined understanding of TNO dynamics. Moreover, these assumptions also represent a limitation for related studies that are not specifically focused on TNOs but use the model of Galactic tides~\citep[see e.g.][]{Veras-Evans_2013}. In addition, this circular-orbit assumption leads to biased estimates of stellar encounter rates~\citep[see][]{MartinezBarbosa-Jilkov-PortegiesZwart-Brown_2017}.

In this context, we revisit the modelling of Galactic tides and their evolution in view of the modern, more complete picture of the Galaxy. A general expression of Galactic tides, free from previous assumptions, would also allow us to explore more diverse trajectories in the Galaxy, including those of extrasolar planetary systems.

In Sect.~\ref{sec:GM}, we present the Galactic model that we used to integrate the Sun's trajectory through the Milky Way. The model is based on the work by~\cite{Khalil_2025}, which is extended in this work to a three-dimensional formulation for our purposes. In Sect.~\ref{sec:GT}, we generalise the Galactic tides model of~\cite{Heisler-Tremaine_1986} in order to make it applicable to any stellar trajectory and any Galactic potential. In Sect.~\ref{sec:EGT}, we explore how the Sun's motion influences Galactic tides, including the newly introduced Galactic tide parameters. In Sect.~\ref{sec:Stats}, we carry out a massive backward integration of the Sun's motion over the last gigayear in order to reveal statistical trends in the evolution of the Galactic tides. Finally, we conclude in Sect.~\ref{sec:cls}.

\section{Model of the Galactic potential}\label{sec:GM}
\citet{Khalil_2025} adjusted a fully non-axisymmetric potential of the Milky Way disc to Gaia DR3 (Data Release~3) data. The axisymmetric background and bar model were inspired from~\citet{Portail-Gerhard-Wegg-Ness_2017} and~\citet{Thomas_2023}. Making use of the backward integration method (based on the conservation of the stellar distribution function in phase space as encoded in the Vlasov equation), the shape of the local velocity distribution could be predicted, and a bar pattern speed of 37~${\rm km} \, {\rm s}^{-1} \, {\rm kpc}^{-1}$ was adjusted to the detailed local velocity distribution measured with Gaia around the Sun. This bar-only model, however, clearly did not account for the full map of median galactocentric stellar radial velocities as a function of position in the Galactic plane, also measured by Gaia. Two spiral modes were subsequently fitted to this median galactocentric radial velocity field over a large portion of the Galactic plane. These spiral modes are a high-amplitude two-armed mode and a low-amplitude three-armed mode, respectively, with distinct pattern speeds. Despite being purely dynamical, the adjustment recovered quite well the actual current positions of spiral overdensities identified photometrically within the Milky Way disc. The low pattern speeds of both spiral patterns (13.1~${\rm km} \, {\rm s}^{-1} \, {\rm kpc}^{-1}$ and 16.4~${\rm km} \, {\rm s}^{-1} \, {\rm kpc}^{-1}$) place their co-rotation radii quite far out, perhaps indicating a link with the interaction of the outer Galactic disc with the Sagittarius (Sgr) dwarf galaxy. The model also remarkably reproduces fine structures in phase-space with exquisite details, such as the moving groups of the disc~\citep[see e.g.][]{Famaey2005,Ramos2018,Bernet2022}, both locally and out to large distances from the solar neighbourhood. 

\begin{figure*}
    \sidecaption
    \includegraphics[width=0.705345\columnwidth]{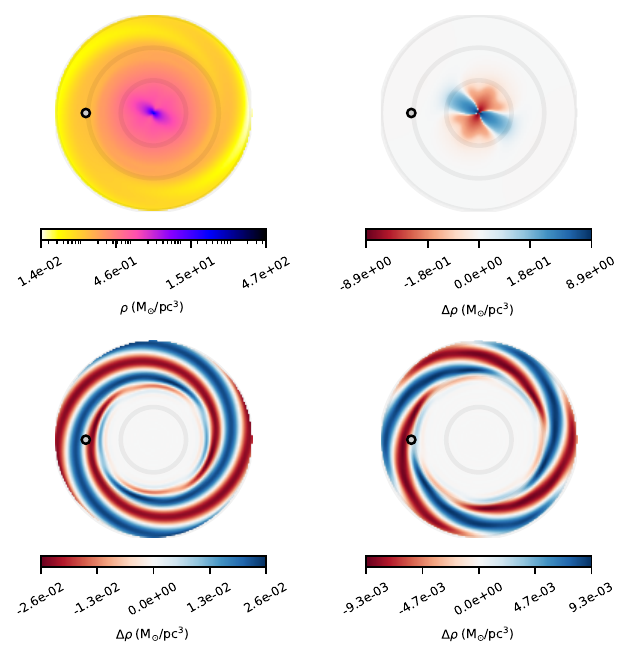}
    \caption{Density of the Galactic model within the Galactic plane ($z=0$). The Sun, indicated with a black circle, is located at $(x,y,z)=(-8.275,0,0.02) \, {\rm kpc}$, and the Galaxy rotates clockwise. \textit{Top-left panel}: total density of the model within the plane, including the axisymmetric part (gas, stars and dark matter), the bar and all spiral modes. The density at the Sun location is 0.108~${\rm M}_\odot \, {\rm pc}^{-3}$. \textit{Top-right panel}: contrast density of the bar. \textit{Bottom panels}: contrast density of the spiral arms for the two-armed mode (\textit{left}) and three-armed mode (\textit{right}).}
    \label{fig:GM:DENS}
\end{figure*}

As per~\citet{Khalil_2025}, the total baryonic mass of the Galaxy (bulge, stellar disc, and gas disc) is set to\footnote{The mass, scale-length, and scale-height are, respectively, $4.32 \times 10^{10} \, {\rm M}_\odot$, 2.4~kpc and 300~pc for the exponential stellar disc profile, and $1.01 \times 10^{10} \, {\rm M}_\odot$, 4.8~kpc and 130~pc for the exponential gas disc profile.} $6 \times 10^{10} \, {\rm M}_\odot$, and the total enclosed mass (baryons and dark matter) within 20~kpc is $2.2 \times 10^{11} \, {\rm M}_\odot$. The dark matter halo is slightly oblate, with an axis ratio $q=0.8$, and has a central constant density core, with the halo power-law slope reaching `$-1$' only at a distance of 3~kpc from the centre. The dark matter density at the solar position is $1.3 \times 10^{-2} \, {\rm M}_\odot \, {\rm pc}^{-3}$. All the in-plane parameters of the bar and spiral arms are defined and given in~\citet[][Table~2]{Khalil_2025}. The bar is extended to three dimensions exactly as in~\citet{Thomas_2023}, while the three-dimensional spiral arms amplitudes~\citep[see][Sect.~3.3, for 2D version]{Khalil_2025}, inspired from the potential-density pairs of~\citet{Cox_2002} and~\citet{Monari-Famaey-Siebert-Grand-Kawata-Boily_2016}, take the form
\begin{equation}
    \phi_{\mathrm{s},m}^{\mathrm{3D}}(R, z, t) = \phi_{\mathrm{s},m}(R, t) \left[ \sech{\left( \frac{W_{\mathrm{s},m} z}{\beta_{\mathrm{s},m}} \right)} \right]^{\beta_{\mathrm{s},m}}\,,
\end{equation}
\[\text{with } W_{\mathrm{s},m} = \frac{m}{R \sin(p_{\mathrm{s},m})} \text{ and } \beta_{\mathrm{s},m} = W_{\mathrm{s},m} h_{\mathrm{s},m} (1 + 0.4 W_{\mathrm{s},m} h_{\mathrm{s},m}).\]
All spiral parameters are as in~\citet[][Table~2]{Khalil_2025}, apart from (i)~a slightly increased spiral scale-height $h_{\mathrm{s},m}=260$~pc, to avoid a too high density contrast out of the plane, and (ii)~a smaller outer radial cut-off of the potential, much inside the outer Lindblad resonance of each spiral mode (cut-off at 12.75~kpc and 11.6~kpc for the two-~and three-armed mode, respectively). The latter does not mean that there is no spiral arm beyond those radii, but that the associated density contrast is not well captured by our simple parametrisation beyond those radii. This does not affect any of the results presented in this paper. At these large radii, the Galactic potential is also known to be non-plane symmetric, which we do not take into account in our modelling. The radial cut-off on the potential is not a Heaviside function but is of the form $H_{m}(R) = (1+\tanh{((R-R_{\rm cut})/\Delta)})/2$, with $\Delta \sim 500 \, {\rm pc}$. Finally, we also set a vertical cut-off on the spiral arms density at $z=130 \, {\rm pc}$. Such a three-dimensional non-axisymmetric Galactic potential can straightforwardly be generated with the {\tt AGAMA}\footnote{\url{https://github.com/GalacticDynamics-Oxford/Agama/blob/master/py/example_mw_potential_khalil25.py}} software~\citep{Vasiliev_2019}. The total density of the model within the Galactic plane is presented on Fig.~\ref{fig:GM:DENS}, together with the contrast density of each non-axisymmetric mode.

In the following, we consider three cases in order to understand the influence of the non-axisymmetric components of the Galactic potential on the tides: the axisymmetric part alone, the axisymmetric part combined with the bar potential, and the full potential, including the two spiral-arm modes. For the sake of simplicity, we later treat these three cases, with a slight abuse of terminology, as three distinct Galactic models.

The axisymmetric part of the potential has no analytical expression; it is generated by \texttt{AGAMA} with the two solvers `CylSpline' (for disc components) and `Multipole' (for spheroidal components) from the analytical axisymmetric densities~\citep{Vasiliev_2019}. As the non-axisymmetric components of the potential of~\cite{Khalil_2025} have analytical expressions, they are directly computed in our code.

\section{Galactic tides}\label{sec:GT}
In this section, we derive expressions for the Galactic tide parameters, which quantify the strength of the Galactic perturbations experienced by a small body orbiting the Sun. Unlike previous works, we use self-consistent equations that are valid for any trajectory of the Sun (or any other star) within the Galaxy and for any Galactic potential, independently of any symmetry considerations.

\begin{figure}
    \includegraphics[width=1.0\columnwidth]{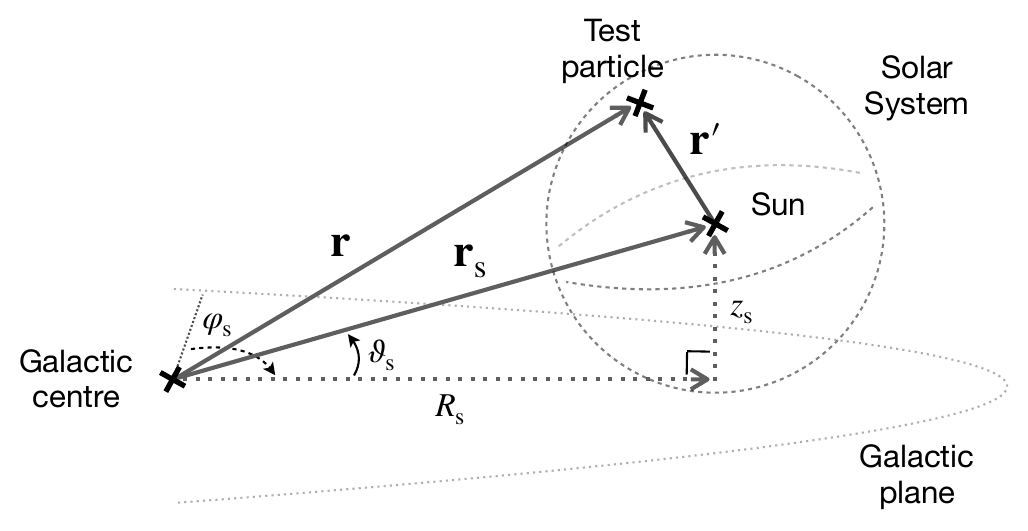}
    \caption{Schematic representation of the Solar System's position relative to the Milky Way's Galactic centre and Galactic plane.}
    \label{fig:GT:triangle}
\end{figure}

We consider a test particle orbiting the Sun and the Sun orbiting the Galaxy. The Sun's motion is not necessarily restricted to the Galactic plane, and there is no imposed constraint on the global shape of its orbit. We denote by $\mathbf{r}_\mathrm{s}$, $\mathbf{r}$, and $\mathbf{r}^\prime$ the radial position vectors of the Sun, of the test particle with respect to the Galactic centre, and of the test particle with respect to the Sun, respectively (see Fig.~\ref{fig:GT:triangle}). It follows that we have the following closed relation:
\begin{equation}
    \label{eq:GT:triangle}
    \mathbf{r}=\mathbf{r}_\mathrm{s}+\mathbf{r}^\prime\,.
\end{equation}

Now, we consider two different right-handed orthonormal frames, one centred at the Galactic centre and the other centred to the Sun. The first corresponds to the inertial fixed galactocentric frame, denoted $\mathcal{E}=(\hat{\mathbf{x}},\hat{\mathbf{y}},\hat{\mathbf{z}})$, such that at $t=0$, the $\hat{x}$-axis points toward the Sun; the $\hat{z}$-axis points toward the north Galactic pole. The second frame is non-inertial and is denoted $\mathcal{E}^\prime=(\hat{\mathbf{x}}^\prime,\hat{\mathbf{y}}^\prime,\hat{\mathbf{z}}^\prime)$; in this frame, the $\hat{x}^\prime$-axis points toward the Galactic centre at all times $t$, the $\hat{y}^\prime$-axis is parallel to the Galactic plane and $\hat{z}^\prime$-axis points in the same hemisphere as $\hat{z}$-axis at $t=0$. We define $\varphi_\mathrm{s}$ and $\vartheta_\mathrm{s}$ as the Sun's azimuthal angle in the Galactic plane and its elevation angle relative to the plane, respectively (see Fig.~\ref{fig:GT:triangle}). The change-of-basis matrix from $\mathcal{E}^\prime$ to $\mathcal{E}$ is then
\begin{equation}
    \label{eq:GT:FtoR}
    \begin{aligned}
         \mathcal{P}_{{\mathcal{E}}^\prime}^\mathcal{E}=\mathcal{R}_{2}(\vartheta_\mathrm{s}(t))\mathcal{R}_{3}(\varphi_\mathrm{s}(t)+\pi)\,,
    \end{aligned}
\end{equation}
where
\begin{equation}
    \begin{aligned}
        \mathcal{R}_{2}(\vartheta_\mathrm{s})=\begin{pmatrix}\cos{\vartheta_\mathrm{s}} & 0 & -\sin{\vartheta_\mathrm{s}} \\ 0 & 1 & 0 \\ \sin{\vartheta_\mathrm{s}} & 0 & \cos{\vartheta_\mathrm{s}}\end{pmatrix}
    \end{aligned}
\end{equation}
and
\begin{equation}
    \begin{aligned}
        \mathcal{R}_{3}(\varphi_\mathrm{s}+\pi)=\begin{pmatrix}-\cos{\varphi_\mathrm{s}} & -\sin{\varphi_\mathrm{s}} & 0 \\ \sin{\varphi_\mathrm{s}} & -\cos{\varphi_\mathrm{s}} & 0 \\ 0 & 0 & 1\end{pmatrix}\,.
    \end{aligned}
\end{equation}

We denote by $\mathbf{r}=(x,y,z)$ and $\mathbf{r}_\mathrm{s}=(x_\mathrm{s}, y_\mathrm{s}, z_\mathrm{s})$ the coordinates of the test particle and of the Sun, respectively, in the $\mathcal{E}$ coordinate system, and by $\mathbf{r}^\prime=(x^\prime,y^\prime,z^\prime)$ the coordinates of the test particle in the $\mathcal{E}^\prime$ coordinate system. The Sun's position is independent of the test particle's position (i.e. $\partial_{(x,y,z)}(x_\mathrm{s}, y_\mathrm{s}, z_\mathrm{s})=0_{3,3}$). So, the Jacobian matrix of the coordinate transformation for the test particle is
\begin{equation}
    \label{eq:GT:diffxyz}
    \begin{aligned}
        \frac{\partial(x,y,z)}{\partial(x^\prime,y^\prime,z^\prime)}=-\begin{pmatrix}x_\mathrm{s}/r_\mathrm{s} & -y_\mathrm{s}/R_\mathrm{s} & z_\mathrm{s}x_\mathrm{s}/(r_\mathrm{s}R_\mathrm{s})\\ y_\mathrm{s}/r_\mathrm{s} & x_\mathrm{s}/R_\mathrm{s} & z_\mathrm{s}y_\mathrm{s}/(r_\mathrm{s}R_\mathrm{s})\\ z_\mathrm{s}/r_\mathrm{s} & 0 & -R_\mathrm{s}/r_\mathrm{s}\end{pmatrix}\,,
    \end{aligned}
\end{equation}
where $R_\mathrm{s}=\sqrt{x_\mathrm{s}^2+y_\mathrm{s}^2}$ and $r_\mathrm{s}=\sqrt{R_\mathrm{s}^2+z_\mathrm{s}^2}$.

According to Newton's Second Law, the force per unit mass acting on the test particle orbiting the Sun under the influence of the Galaxy is:
\begin{equation}
    \label{eq:GT:PFD:1}
    \ddot{\mathbf{r}}=-\frac{\mu}{{r^\prime}^3}\mathbf{r}^\prime-\boldsymbol{\nabla}\Phi\,,
\end{equation}
where $\mu$ is the gravitational parameter of the Sun and $\Phi$ is the Galactic potential. Considering that $\|\mathbf{r}^\prime\|\ll\|\mathbf{r}_\mathrm{s}\|$, the gradient of the Galactic potential for the test particle near the Sun can be approximated using a first-order Taylor expansion around the Sun:
\begin{equation}
    \label{eq:GT:gradPhi:1}
    \begin{aligned}
        \boldsymbol{\nabla}\Phi=\left.\boldsymbol{\nabla}\Phi\right|_{\mathbf{r}_\mathrm{s}}+\left.\boldsymbol{\nabla}\boldsymbol{\nabla}\Phi\right|_{\mathbf{r}_\mathrm{s}}\delta\mathbf{r}+\mathcal{O}(\|\delta\mathbf{r}\|^2)\,,
    \end{aligned}
\end{equation}
where
\begin{equation}
    \label{eq:GT:gradPhi:1:comp1}
    \begin{aligned}
        \left.\boldsymbol{\nabla}\Phi\right|_{\mathbf{r}_\mathrm{s}}=-\ddot{\mathbf{r}}_\mathrm{s}
    \end{aligned}
\end{equation}
is the acceleration of the Sun within the Milky Way,
\begin{equation}
    \label{eq:GT:gradPhi:1:comp2}
    \begin{aligned}
        \delta\mathbf{r}&=\mathbf{r}-\mathbf{r}_\mathrm{s}=\mathbf{r}^{\prime}
    \end{aligned}
\end{equation}
is the small displacement of the test particle relative to the Sun, and
\begin{equation}
    \label{eq:GT:gradPhi:1:comp3}
    \begin{aligned}
        \left.\boldsymbol{\nabla}\boldsymbol{\nabla}\Phi\right|_{\mathbf{r}_\mathrm{s}}=\begin{pmatrix}\partial^2_{x^\prime x^\prime}\Phi & \partial^2_{x^\prime y^\prime}\Phi & \partial^2_{x^\prime z^\prime}\Phi\\
        \partial^2_{y^\prime x^\prime}\Phi & \partial^2_{y^\prime y^\prime}\Phi & \partial^2_{y^\prime z^\prime}\Phi\\
        \partial^2_{z^\prime x^\prime}\Phi & \partial^2_{z^\prime y^\prime}\Phi & \partial^2_{z^\prime z^\prime}\Phi\end{pmatrix}_{x_\mathrm{s},y_\mathrm{s},z_\mathrm{s}}
    \end{aligned}
\end{equation}
is the Hessian matrix of the Galactic potential evaluated at the Sun's instantaneous position $\mathbf{r}_\mathrm{s}$.

Using the Leibniz chain rule (and Schwarz's theorem), together with Eq.~\eqref{eq:GT:diffxyz} and the definitions of the polar coordinates $R$ and $\varphi$, and then combining Eqs.~\eqref{eq:GT:triangle},~\eqref{eq:GT:PFD:1}, and~\eqref{eq:GT:gradPhi:1}, the equation of motion of the test particle with respect to the Sun is
\begin{equation}
    \label{eq:GT:PFD:2}
    \begin{aligned}
        \ddot{\mathbf{r}}^\prime=&-\frac{\mu}{{r^\prime}^3}\mathbf{r}^\prime\\
        &-\left(G_1 x^\prime+G_4 y^\prime+G_5 z^\prime\right)\hat{\mathbf{x}}^\prime\\
        &-\left(G_4 x^\prime+G_2 y^\prime+G_6 z^\prime\right)\hat{\mathbf{y}}^\prime\\
        &-\left(G_5 x^\prime+G_6 y^\prime+G_3 z^\prime\right)\hat{\mathbf{z}}^\prime\,,
    \end{aligned}
\end{equation}
where $G_{i}$ ($i=1,\ldots,6$) are the Galactic tide parameters, defined as
\begin{equation}
    \label{eq:GT:Gt}
    \left\{
    \begin{aligned}
        G_1=&\left(\frac{R^2}{r^2}\partial^2_{RR}\Phi+2\frac{Rz}{r^2}\partial^2_{Rz}\Phi+\frac{z^2}{r^2}\partial^2_{zz}\Phi\right)_{R_\mathrm{s},\varphi_\mathrm{s},z_\mathrm{s}}\,,\\
        G_2=&\left(\frac{1}{R}\partial_{R}\Phi+\frac{1}{R^2}\partial^2_{\varphi\varphi}\Phi\right)_{R_\mathrm{s},\varphi_\mathrm{s},z_\mathrm{s}}\,,\\
        G_3=&\left(\frac{z^2}{r^2}\partial^2_{RR}\Phi-2\frac{Rz}{r^2}\partial^2_{Rz}\Phi+\frac{R^2}{r^2}\partial^2_{zz}\Phi\right)_{R_\mathrm{s},\varphi_\mathrm{s},z_\mathrm{s}}\,,\\
        G_4=&\left(-\frac{1}{rR}\partial_{\varphi}\Phi+\frac{1}{r}\partial^2_{R\varphi}\Phi+\frac{z}{rR}\partial^2_{\varphi z}\Phi\right)_{R_\mathrm{s},\varphi_\mathrm{s},z_\mathrm{s}}\,,\\
        G_5=&\left(\frac{Rz}{r^2}\partial^2_{RR}\Phi+\frac{z^2-R^2}{r^2}\partial^2_{Rz}\Phi-\frac{Rz}{r^2}\partial^2_{zz}\Phi\right)_{R_\mathrm{s},\varphi_\mathrm{s},z_\mathrm{s}}\,,\\
        G_6=&\left(-\frac{z}{rR^2}\partial_{\varphi}\Phi+\frac{z}{rR}\partial^2_{R\varphi}\Phi-\frac{1}{r}\partial^2_{\varphi z}\Phi\right)_{R_\mathrm{s},\varphi_\mathrm{s},z_\mathrm{s}}\,.
    \end{aligned}
    \right.
\end{equation}
The Galactic component of the force in Eq.~\eqref{eq:GT:PFD:2} is derived from the potential:
\begin{equation}
    \label{eq:GT:potGF}
    \begin{aligned}
        \widetilde{\Phi}(x^\prime,y^\prime,z^\prime)=&\quad\frac{G_1}{2}{x^{\prime}}^2+\frac{G_2}{2}{y^{\prime}}^2+\frac{G_3}{2}{z^{\prime}}^2\\
        &+G_4x^{\prime}y^{\prime}+G_5x^{\prime}z^{\prime}+G_6y^{\prime}z^{\prime}\,.
    \end{aligned}
\end{equation}
Finally, we note that
\begin{equation}
    \label{eq:GT:lp}
    \begin{aligned}
        \Delta^\prime\widetilde{\Phi}=\sum_{i=1}^3 G_i=\left.\Delta\Phi\right|_{\mathbf{r}_\mathrm{s}}=4\pi\mathcal{G}\rho_0\,,
    \end{aligned}
\end{equation}
which corresponds to Poisson's equation evaluated in the solar neighbourhood ($\mathcal{G}$ is the gravitational constant). Thus, $\rho_0$ is the local mass density of the Galaxy in the solar neighbourhood.\footnote{We stress that the mass density $\rho_0$ here is representative of the mean field. Hence it is the coarse-grained smooth stellar density (with also gas and dark matter) on the scale of $\sim 20$~pc, not the fine-grained discrete small-scale density.}

The expressions in Eq.~\eqref{eq:GT:Gt} are valid for any Galactic potential, and any trajectory of the Sun. If the potential is axisymmetric (i.e. $\Phi=\Phi(R,z)$) and symmetric with respect to the plane~$z=0$ (i.e. $\Phi(R,-z)=\Phi(R,z)$) and the Sun follows a circular orbit in the Galactic plane (i.e. $\dot{R}_\mathrm{s}=0$ and $z_\mathrm{s}=0$), we retrieve the well-known form of the Galactic tide parameters derived by~\cite{Heisler-Tremaine_1986}, namely:
\begin{equation}
    \label{eq:GT:Gtaxi}
    \left\{
    \begin{aligned}
        G_1=&\left.\partial^2_{RR}\Phi\right|_{R_\mathrm{s},\varphi_\mathrm{s},z_\mathrm{s}}\\
        G_2=&\frac{1}{R_\mathrm{s}}\left.\partial_{R}\Phi\right|_{R_\mathrm{s},\varphi_\mathrm{s},z_\mathrm{s}}\\
        G_3=&\left.\partial^2_{zz}\Phi\right|_{R_\mathrm{s},\varphi_\mathrm{s},z_\mathrm{s}}
    \end{aligned}
    \qquad\,,\qquad G_4=G_5=G_6=0\,.
    \right.
\end{equation}

We note that in previous papers, it was common to consider only two components for the Galactic tides: the `normal' component characterised by $G_3$, and the `radial' component characterised by $G_1$ and $G_2$ \citep[see, e.g.][]{Matese-Whitmire_1996}. To avoid any confusion, we abandon this terminology and treat all the Galactic tide parameters in Eq.~\eqref{eq:GT:Gt} independently.

\section{Evolution of Galactic tides}\label{sec:EGT}
In this section, we show how the values of the Galactic tide parameters defined in Sect.~\ref{sec:GT} are qualitatively affected by the Sun's trajectory in the Galactic potential described in Sect.~\ref{sec:GM}. First, in Sect.~\ref{sec:EGT1}, we revisit the approximations made in previous works and examine the dominant components of the Galactic potential. Second, in Sect.~\ref{sec:EGT2}, we avoid all assumptions. This allows us to qualitatively understand the roles played by different properties of the potential in the evolution of Galactic tides.

\subsection{Toy model under standard assumptions}\label{sec:EGT1}
In the axisymmetric limit, assuming that the Sun's motion is circular and confined to the Galactic plane, the Galactic tide parameters in Eq.~\eqref{eq:GT:Gtaxi} can be expressed in terms of the standard Oort constants, $A$ and $B$, as
\begin{equation}
    \label{eq:EGT:GTaxi}
    \left\{
    \begin{aligned}
        G_1&=-(A-B)(3A+B)\,,\\
        G_2&=(A-B)^2\,,\\
        G_3&=4\pi\mathcal{G}\rho_0-2(B^2-A^2)\,;
    \end{aligned}
    \right.
\end{equation}
see~\cite{Heisler-Tremaine_1986} for details. Current measurements indicate that $A\approx-B$~\citep{Gunn-Knapp-Tremaine_1979,Olling-Dehnen_2003,Li-Zhao-Yang_2019,Guo-Qi_2023}, which justifies the approximation 
\begin{equation}
    \label{eq:EGT:G1eq-G2}
    \begin{aligned}
        G_2\approx -G_1\,
    \end{aligned}
\end{equation}
adopted in many studies (see e.g.~\citealp{Heisler-Tremaine_1986,Levison-Dones-Duncan_2001,Fouchard_2004,Gardner-Nurm-Flynn-Mikkola_2011,Saillenfest-Fouchard-Ito-Higuchi_2019}). Equation~\eqref{eq:EGT:GTaxi} with approximation in Eq.~\eqref{eq:EGT:G1eq-G2} is what we define as our toy model.

\begin{figure*}
    \includegraphics[width=1\columnwidth]{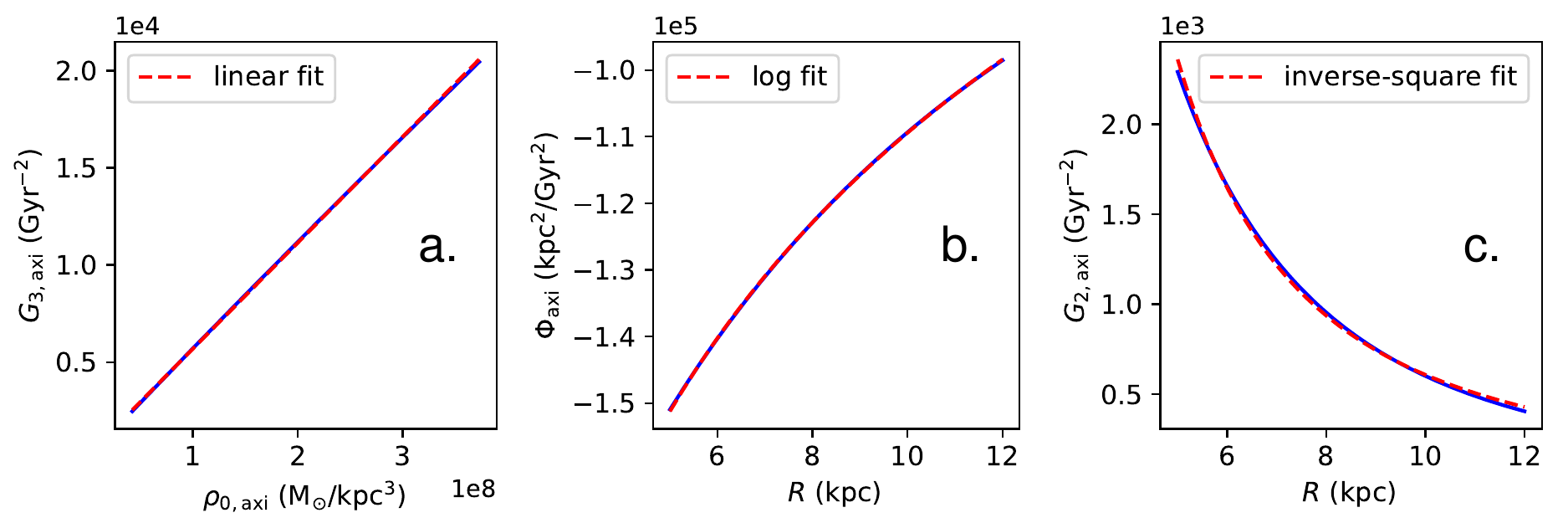}
    \caption{\textit{Panel~a} : Linear fit (red dashed line) of $G_3$ at $z = 0$ (blue line), considering only the axisymmetric part of the Galactic potential presented in Sect.~\ref{sec:GM} (axisymmetric case). \textit{Panel~b} : Logarithmic fit (red dashed line) of the axisymmetric Galactic potential (defined in Sect.~\ref{sec:GM}) evaluated at~$z = 0$ (blue line). \textit{Panel~c} : Inverse-square fit (red dashed line) of $G_2$ at $z = 0$ in the axisymmetric case, $\Phi=\Phi_{\text{axi}}$ (blue line).}
    \label{fig:EGT:combinedfits}
\end{figure*}

From Eqs.~\eqref{eq:GT:lp} and~\eqref{eq:EGT:G1eq-G2}, it follows that
\begin{equation}
    \label{eq:EGT:G3proptoRho}
    \begin{aligned}
        G_3\propto\rho_0\,,
    \end{aligned}
\end{equation}
indicating that variations in $\rho_0$ directly reflect changes in $G_3$ (see Fig.~\ref{fig:EGT:combinedfits}a). We note that in the case of a strict equality $G_2=-G_1$, the Galactic potential within the toy model is reduced to a logarithmic potential\footnote{A logarithmic potential provides a constant velocity curve, which is --- at first approximation --- valid at the Sun position~\citep[see, e.g.][]{Eilers-Hogg-Rix-Ness_2019}.} $\Phi\propto\log{R}$, provided that\footnote{Assuming $R_\mathrm{s}\gg z_\mathrm{s}$ is a reasonable approximation for the Sun; see later in Sect.~\ref{sec:Stats:results}.} $R_\mathrm{s}\gg z_\mathrm{s}$. Figure~\ref{fig:EGT:combinedfits}b shows that the axisymmetric background of the Galactic potential presented in Sect.~\ref{sec:GM} indeed behaves like a logarithmic function. In this case, $G_2\propto 1/R^2$ (see Fig.~\ref{fig:EGT:combinedfits}c), so $G_2$ shares the same periodicity as $R$ and increases as $R$ decreases (and decreases as $R$ increases).

\begin{figure}
    \includegraphics[width=0.7\columnwidth]{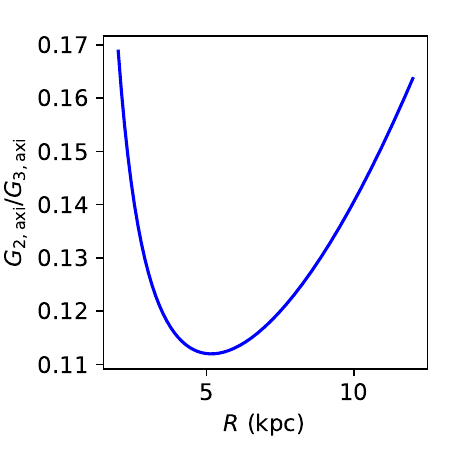}
    \caption{Ratio of $G_2/G_3$ at $z = 0$ in the axisymmetric case (i.e. $\Phi=\Phi_{\text{axi}}$; see Fig.~\ref{fig:EGT:combinedfits}b).}
    \label{fig:EGT:ratioG2onG3}
\end{figure}

Figure~\ref{fig:EGT:ratioG2onG3} shows the ratio $G_2/G_3$ in the axisymmetric case (i.e. considering only the axisymmetric component of the full Galactic potential presented in Sect.~\ref{sec:GM}). As visible in Fig.~\ref{fig:EGT:ratioG2onG3}, the common approximation made in previous papers to neglect $G_1$ and $G_2$ as compared to $G_3$ is questionable, even when the potential is restricted to the axisymmetric component of the Galaxy. This was already stressed by~\cite{Fouchard_2004} for numerical studies. Nevertheless, it remains a useful approximation for analytical studies, as it reduces the number of degrees of freedom in the Hamiltonian and makes the problem integrable (see e.g.~\citealp{Heisler-Tremaine_1986,Breiter-Dybczynski-Elipe_1996,Saillenfest-Fouchard-Ito-Higuchi_2019}).

\begin{figure}
    \includegraphics[width=1.\columnwidth]{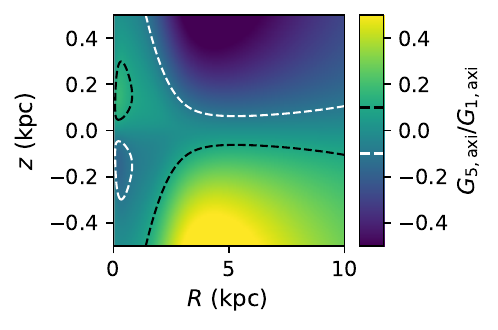}
    \caption{Ratio of $G_5/G_1$ in the axisymmetric case (i.e. $\Phi=\Phi_{\text{axi}}$; see Fig.~\ref{fig:EGT:combinedfits}b). The white and black curves correspond to values of about $-10\%$ and $+10\%$, respectively; they therefore indicate the region where the approximation~$|G_1| \gg |G_5|$ holds.}
    \label{fig:EGT:ratioG5onG1}
\end{figure}

For an axisymmetric potential, one gets $G_4=G_6=0$ from Eq.~\eqref{eq:GT:Gt}. Regarding $G_5$, it cancels when $z_\mathrm{s}=0$, and the potential is symmetric with respect to the Galactic plane ($z=0$). Figure~\ref{fig:EGT:ratioG5onG1} shows how the ratio $G_5/G_1$ evolves as the Sun moves away from the Galactic plane. This indicates that, depending on the Sun's trajectory in the $(R,z)$-plane and on the choice of axisymmetric potential, assuming $G_5=0$ as in Eq.~\eqref{eq:GT:Gtaxi} may have a minor to significant impact on the results. Strictly speaking, the classic formulas in Eq.~\eqref{eq:GT:Gtaxi} are only valid for a very restrictive set of assumptions; one should not use them for more general Galactic trajectories (as done by e.g.~\citealp{Gardner-Nurm-Flynn-Mikkola_2011}). In Appendix~\ref{an:add-fig_comp}, we explore this ratio for the axisymmetric model of~\cite{Gardner-Nurm-Flynn-Mikkola_2011} and examine which region of the $(R,z)$-plane their Sun's orbit explores; we show that the parameter $G_5$, ignored in their study, would actually reach up to 20\% of the value of $G_1$ in their model.

\subsection{Realistic non-axisymmetric model}\label{sec:EGT2}
From now on, $G_{\text{i}\text{---}\text{j}}$ denote the set of all $G_k$ such that $k\in\mathbb{N}$ and $\text{i}\le k\le\text{j}$. For instance, $G_{2\text{---}4}$ means the set $\left\{G_2,\,G_3,\,G_4\right\}$.

\begin{figure*}
    \includegraphics[width=0.92\columnwidth]{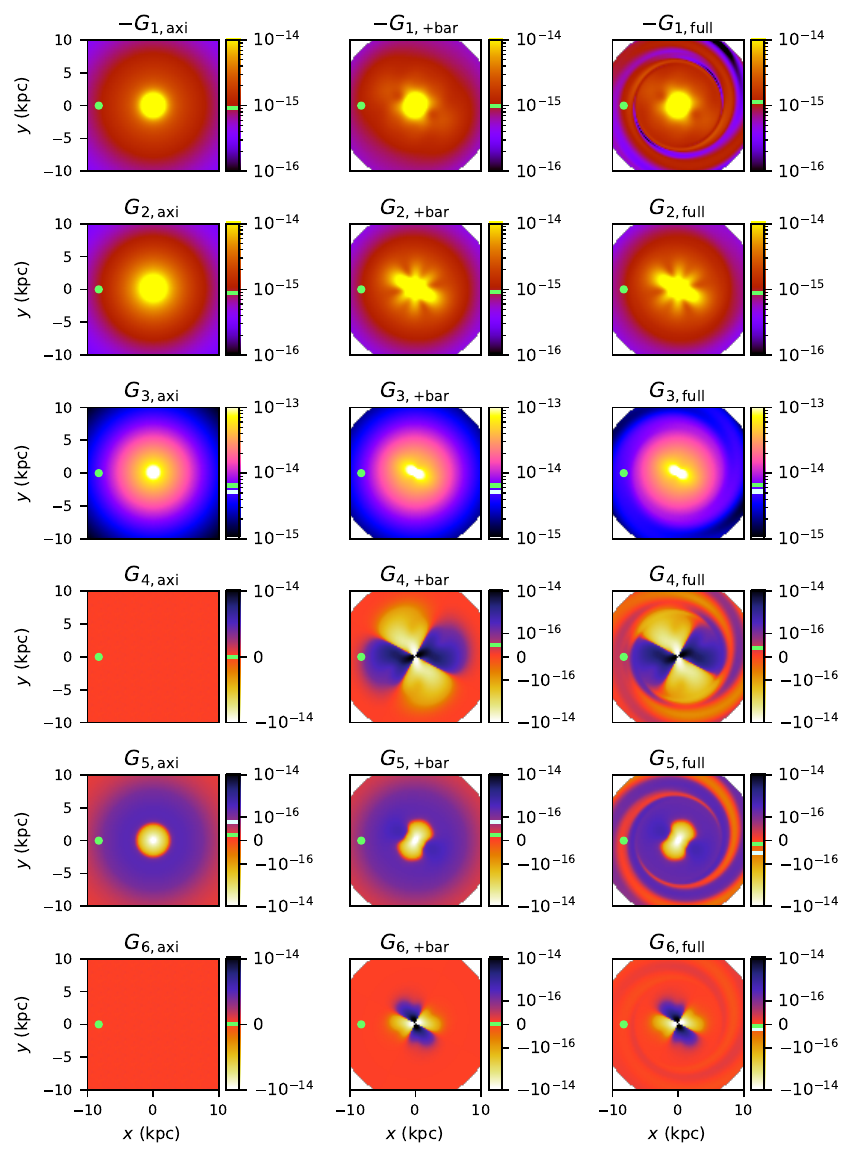}
    \caption{Present-day snapshot of the Galactic tidal fields at $z=70$~pc, shown for the axisymmetric background of the Galactic potential (\textit{left column}), for the axisymmetric background plus the Galactic bar (\textit{middle column}), and for the full model (\textit{right column}). Colour units are per year squared (i.e. yr$^{-2}$). The green dot marks the Sun at $(x,y,z)=(-8.275,0,0.02) \, {\rm kpc}$; the green line on the colour bar indicates the Sun's current value; the light blue-grey line indicates the Sun’s value for $z_\mathrm{s}=70$~pc instead of $20$~pc (in contrast to the green line).}
    \label{fig:EGT:colormapG}
\end{figure*}

Figure~\ref{fig:EGT:colormapG} shows a present-day snapshot of the Galactic tidal fields ($G_{1\text{---}6}$) for the three Galactic models defined in Sect.~\ref{sec:GM}: the axisymmetric background of the Galactic potential, the axisymmetric background plus the Galactic bar, and the full model. More specifically, these snapshots in Fig.~\ref{fig:EGT:colormapG} consist of colour maps representing $(x,y)$-plane slices of the Galactic tidal fields at $z=70$~pc. This value of $z$ roughly corresponds to the median vertical height reached by the Sun\footnote{Here, we do not refer to the median vertical position of the Sun.} (see later in Sect.~\ref{sec:Stats:results}); we chose a non-zero value of $z$ in order to highlight the variations in the parameters $G_5$ and $G_6$ (see Eq.~\eqref{eq:GT:Gt} or Appendix~\ref{an:add-fig_color}). Appendix~\ref{an:add-fig_color} contains two additional figures reproducing Fig.~\ref{fig:EGT:colormapG} for $z=0$ and $z=20$~pc.

For the axisymmetric model, we see that the trends discussed in the previous section appear to be preserved, meaning that our axisymmetric model behaves like the idealised case, namely $G_1\approx -G_2$, $|G_{1\text{---}2}| < |G_3|$, $|G_3| \gg |G_5|$ and $G_4=G_6=0$. For the barred model, the influence of the bar is clearly visible in the central regions of the Galaxy, and, compared to the axisymmetric model, no drastic change is noticeable in the Galactic neighbourhood of the Sun, except for $G_4$ which may be significantly non-zero. In contrast, the addition of the spiral arms strongly affects the values of $G_1$, $G_4$ and $G_5$, with slight effects also visible in $G_3$ and $G_6$. Qualitatively, we observe that in the present-day Galactic region around the Sun, the parameters $G_2$, $G_3$ and $G_6$ are not affected so much by the non-axisymmetric features of the model used. Finally, we observe that in all cases we have $|G_3|\gg |G_{4\text{---}6}|$ and $|G_{1\text{---}2}| < |G_3|$ for the present-day Galactic location of the Sun (see Fig.~\ref{fig:EGT:colormapG}).

\begin{figure}
    \includegraphics[width=1\columnwidth]{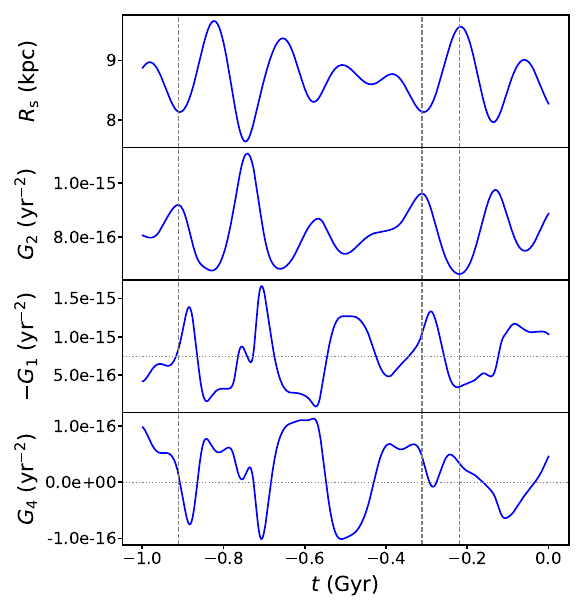}
    \caption{Time evolution over the past $1$~Gyr of $R_\mathrm{s}$, $G_2$, $-G_1$, and $G_4$ using the full Galactic model. All quantities are defined in Sect.~\ref{sec:GT}. The grey vertical dotted lines are included to guide the eye (see text).}
    \label{fig:EGT:signalex1}
\end{figure}

\begin{figure}
    \includegraphics[width=1\columnwidth]{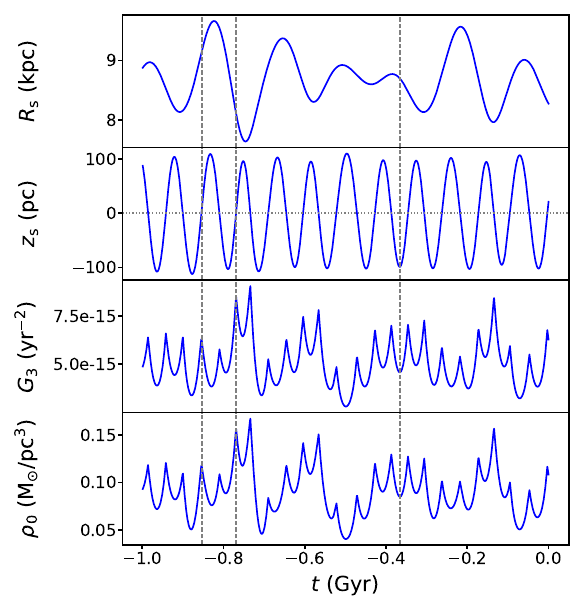}
    \caption{Same as Fig.~\ref{fig:EGT:signalex1} but showing $R_\mathrm{s}$, $z_\mathrm{s}$, $G_3$, and $\rho_0$.}
    \label{fig:EGT:signalex2}
\end{figure}

\begin{figure}
    \includegraphics[width=1\columnwidth]{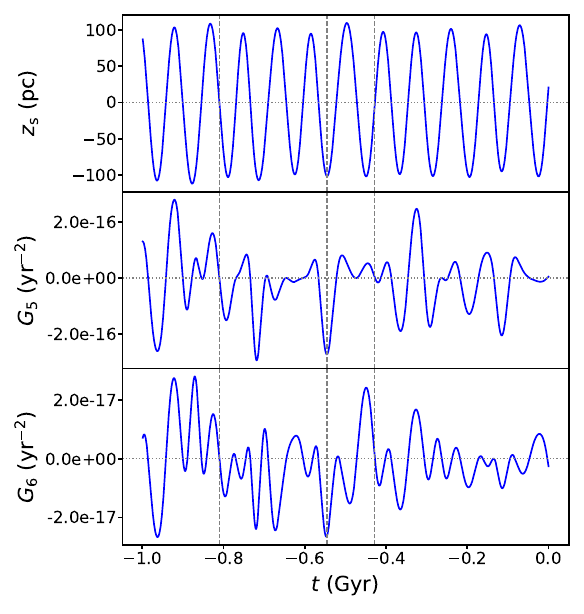}
    \caption{Same as Fig.~\ref{fig:EGT:signalex1} but showing $z_\mathrm{s}$, $G_5$, and $G_6$.}
    \label{fig:EGT:signalex3}
\end{figure}

In order to visualise how the Galactic tide parameters vary as a function of $R_\mathrm{s}$ and $z_\mathrm{s}$ in the full Galactic model, Figs.~\ref{fig:EGT:signalex1}, \ref{fig:EGT:signalex2} and~\ref{fig:EGT:signalex3} show the time evolution of the Sun's coordinates ($R_\mathrm{s}$ and $z_\mathrm{s}$), Galactic tides parameters, and local Galactic mass density. The solar trajectory used here corresponds to the nominal orbit of the Sun according to the statistics explored in Sect.~\ref{sec:Stats}. For now, we only use this trajectory for illustration purposes, in order to qualitatively identify the correlations between variations in the Sun's position and variations in~$G_{1\text{---}6}$; the quantitative exploration is presented in Sect.~\ref{sec:Stats:results}.

\begin{figure*}
    \includegraphics[width=1.0\columnwidth]{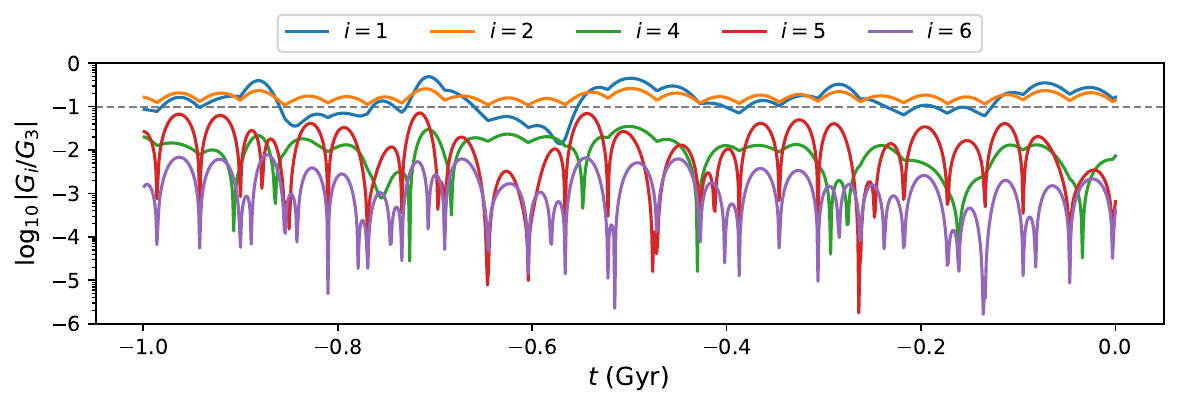}
    \caption{Time evolution of the Galactic tide parameter ratios $\left|G_i/G_3\right|$ (for $i=1,2,4,5,6$) along the Sun’s nominal orbit, derived from the full Galactic model.}
    \label{fig:EGT:Gi_G3}
\end{figure*}

We observe that the variation of $R_\mathrm{s}$ drives that of $G_2$: The maxima of $R_\mathrm{s}$ correspond to the minima of $G_2$, and vice versa (see e.g. the grey, dashed vertical lines in Fig.~\ref{fig:EGT:signalex1}). The variation of $-G_1$ shares some similarities with that of $G_2$, but it is affected by the non-axisymmetry of the potential (primarily due to spiral arms; see Fig.~\ref{fig:EGT:colormapG}), leading to perceptible differences between the two curves --- in the axisymmetric case, $-G_1$ and $G_2$ would behave the same way. The variation of $G_3$ is primarily driven by the rapid variations of $z_\mathrm{s}$, while the slower variations are modulated by $R_\mathrm{s}$: The extrema of $z_\mathrm{s}$ correspond to the minima of $G_3$, whereas the zero-crossings of $z_\mathrm{s}$ correspond to the maxima of $G_3$ (see e.g. the grey, dashed vertical lines in Fig.~\ref{fig:EGT:signalex2}); meanwhile, the slower oscillations of $R_\mathrm{s}$ shape the overall envelope of $G_3$. This is not surprising because $G_3$ is the main vertical component, quantifying the gradient of the vertical force, and most of the mass is concentrated in the Galactic plane; therefore, $G_3$ reaches its maximum in the Galactic plane. Moreover, the mass density increases toward the Galactic centre; consequently, for a fixed $z_\mathrm{s}$, a maximum at smaller $R_\mathrm{s}$ is generally expected to be higher than a maximum at larger $R_\mathrm{s}$. This is even more striking when comparing $\rho_0$ with $G_3$: The $\rho_0$ curve closely resembles that of $G_3$. From Eq.~\eqref{eq:GT:lp}, we deduce that the sum of $G_1$ and $G_2$ is significantly smaller than $G_3$. However, the last important point we need to note here is that $G_1$ and $G_2$ individually might no longer be negligible with respect to $G_3$. This is particularly true for $G_1$. Indeed, Fig.~\ref{fig:EGT:Gi_G3} shows that $-G_1$ can be nearly half of $G_3$. Finally, the curves of $G_4$, $G_5$ and $G_6$ are much harder to interpret. Nevertheless, we observe that the curves $G_4$ and $-G_1$ appear to be approximately mirror images of each other, similar to what is observed for the curves $R_\mathrm{s}$ and $G_2$. We also observe that zero-crossings of $z_\mathrm{s}$ imply zero-crossings of $G_5$ and $G_6$ (see e.g. the grey, dashed vertical lines in Fig.~\ref{fig:EGT:signalex3}), but $G_5=0$ or $G_6=0$ does not imply that $z_\mathrm{s}=0$ (see Fig.~\ref{fig:EGT:signalex3}). Moreover, some of the extrema of $z_\mathrm{s}$ coincide with those of $G_5$ and/or $G_6$; in the axisymmetric case, they coincide with those of $G_5$. Globally, we also note that: $|G_{1\text{---}2}|>|G_{4\text{---}6}|$ (see Fig.~\ref{fig:EGT:Gi_G3}). We verified that the analysis made here remains qualitatively relevant for any Galactic trajectory in the solar neighbourhood.

\section{Statistics derived from the solar galactic orbits}\label{sec:Stats}
As the Sun's initial conditions in the Galaxy are not perfectly known, trajectories can substantially vary from one simulation to another, even within a given Galactic potential. In this section, we sample trajectories in the uncertainty ranges of the Sun's initial conditions and analyse the statistics of the outcomes. The model of the Galactic potential described in Sect.~\ref{sec:GM} has a limited validity over time. Here, we use an optimistic upper bound of $1$~Gyr, which is the duration of all our integrations below.

\subsection{Initial conditions}\label{sec:Stats:orbits}
In order to account for the uncertainties in the present-day coordinates of the Sun, we performed numerical integrations of one million solar trajectories over the past $1$~Gyr. Three sets of integrations were then carried out: one using the Galaxy model described previously in Sect.~\ref{sec:GM}, one using only its axisymmetric component, and one without the spiral arms (i.e., with only the axisymmetric background and the Galactic bar), in order to assess the influence of the different non-axisymmetric structures. 

The initial radial position,
\begin{equation}
    \begin{aligned}
       R_{\mathrm{s},0}=8275\pm100\,\text{pc}\,,
    \end{aligned}
\end{equation}
was taken from~\citet[][Table B.1]{Gravity-Collaboration_2024}, whereas the initial vertical position,
\begin{equation}
    \begin{aligned}
       z_{\mathrm{s},0}=20.8\pm10\,\text{pc}\,,
    \end{aligned}
\end{equation}
was taken from~\cite{Bennett-Bovy_2019}. We note that the value reported by~\cite{Gravity-Collaboration_2024} is $0.1$~kpc larger than that reported by~\cite{GRAVITYCollaboration_2019}, namely $R_{\mathrm{s},0}=8.178 \pm 0.013_{\text{stat.}} \pm 0.022_{\text{sys.}}$~kpc, far exceeding the statistical or systematic uncertainties estimated in~2019. Therefore, it is prudent to assume that the systematic uncertainty of the~2024 measurement could also be of the order of $0.1$~kpc. The same consideration applies to $z_{\mathrm{s},0}$: \cite{Griv-Gedalin-Pietrukowicz-Majaess-Jiang_2021} shows that $z_{\mathrm{s},0}$ varies between~$5$ and $26$~pc, which is far beyond the $0.3$~pc uncertainty mentioned by~\cite{Bennett-Bovy_2019}. Consequently, we adopt $10$~pc as a more conservative estimate of the uncertainty on $z_{\mathrm{s},0}$. The initial solar velocity\footnote{The initial velocity vector is expressed in the fixed cylindrical galactocentric frame.},
\begin{equation}
    \begin{aligned}
       \mathbf{v}_{\mathrm{s},0}=\begin{pmatrix}-11.1^{\pm 1.25} & 250.2^{\pm 3} & 7.25^{\pm 0.6}\end{pmatrix}\,\text{km/s}\,,
    \end{aligned}
\end{equation}
was taken from~\cite{Schonrich-Binney-Dehnen_2010} for the radial and vertical components (summing up the statistical and systematic errors in quadrature), while the tangential component was adjusted to match our adopted value of the initial radial position from the proper motion of Sgr~A*. 

Both the initial positions and velocities were sampled from Gaussian distributions, with individual components drawn independently to construct a sample of one million initial conditions. The same initial-condition sample was used for each Galactic model. The equations of motion of the Sun are given by Eq.~\eqref{eq:GT:gradPhi:1:comp1}.

\subsection{Integrator}
We used the $\mathcal{SABA}_2$ symplectic integrator~\citep{Laskar-Robutel_2001} combined with the isochrone splitting scheme~\citep{Bougakov-Saillenfest-Fouchard_2025}. This approach provides a better trade-off between computational cost and energy conservation than the kinetic splitting scheme in the context of stellar motion in effective gravitational fields (see~\citealp{Bougakov-Saillenfest-Fouchard_2025} for details). Appendix~\ref{an:Perf_int} describes how to set the isochrone splitting parameters in order to achieve better efficiency than the standard Leapfrog-like integrators. We used a timestep of $1$~Myr and integrate over $1$~Gyr, which yields a total number of points of $N=1000$. The relative conservation of energy (when considering an extended Hamiltonian that is conserved over time; see Appendix~\ref{an:Perf_int}) is about~$6\times10^{-8}$ for all integrations. This strict constraint in energy conservation is used here in order to avoid any possible bias in the statistics over 1 million trajectories. The CPU time needed to integrate all $10^6$~trajectories with this level of precision is about $5$~hours\footnote{Time calculated with the full Galactic model; shorter times are expected for the other models.}. The same integrations would take about $32$~hours of CPU time using the kinetic splitting scheme.

\subsection{Statistics on Galactic tide parameters}\label{sec:Stats:results}
\begin{figure}
    \includegraphics[width=1.0\textwidth]{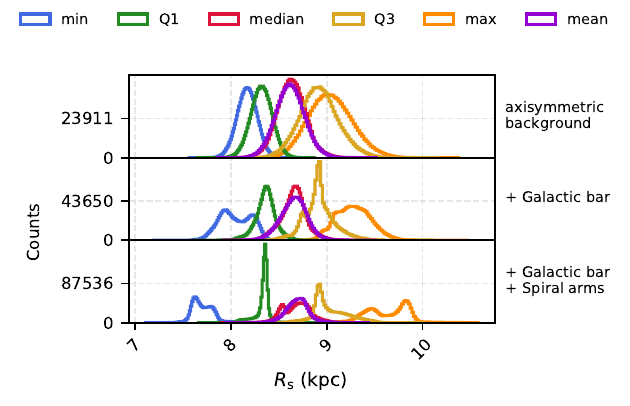}
    \includegraphics[width=1.0\textwidth]{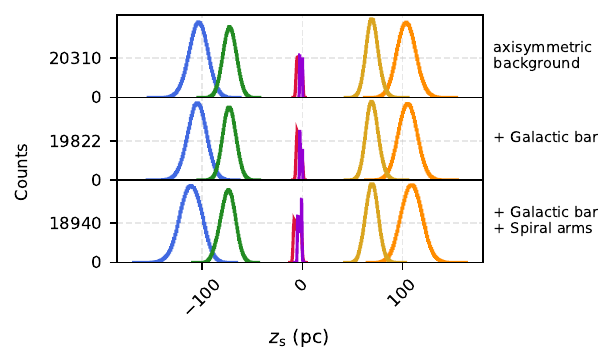}
    \caption{Distributions of trajectory-based summary statistics (minimum, maximum, first three quartiles, and arithmetic mean) of the Sun's galactocentric cylindrical coordinates, shown for the axisymmetric background of the Galactic potential (\textit{top row}), for the axisymmetric background plus the Galactic bar (\textit{middle row}), and for the full model (\textit{bottom row}).}
    \label{fig:Stats:histograms:R_z}
\end{figure}

For each trajectory sampled over $N=1000$ points regularly spaced in time, we compute the minimum, maximum, the first three quartiles (Q1, median, Q3), and the arithmetic mean of: $R_\mathrm{s}$, $z_\mathrm{s}$, $G_{1\text{---}6}$, and $\rho_0$. These quantities are then grouped and plotted as histograms in Figs.~\ref{fig:Stats:histograms:R_z}, \ref{fig:Stats:histograms:G1_G2_G3} and~\ref{fig:Stats:histograms:rho}. Our sampling implies that the Sun spends one quarter of its total dynamical evolution in each of the four intervals: minimum–Q1, Q1–median, median–Q3, and Q3–maximum.

\begin{figure}
    \includegraphics[width=1.0\textwidth]{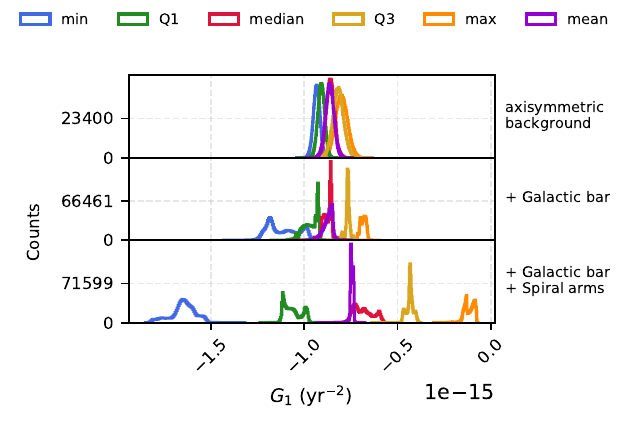}
    \includegraphics[width=1.0\textwidth]{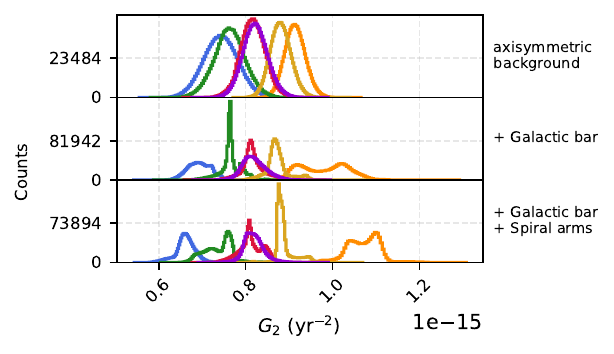}
    \includegraphics[width=1.0\textwidth]{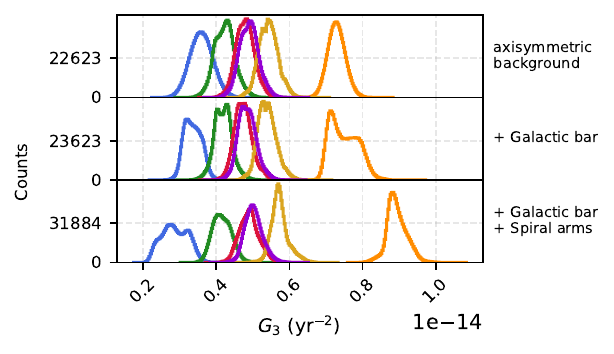}
    \caption{Same as Fig.~\ref{fig:Stats:histograms:R_z} but for trajectory-based summary statistics of Galactic tide parameters.}
    \label{fig:Stats:histograms:G1_G2_G3}
\end{figure}
\addtocounter{figure}{-1}
\begin{figure}
    \includegraphics[width=1.0\textwidth]{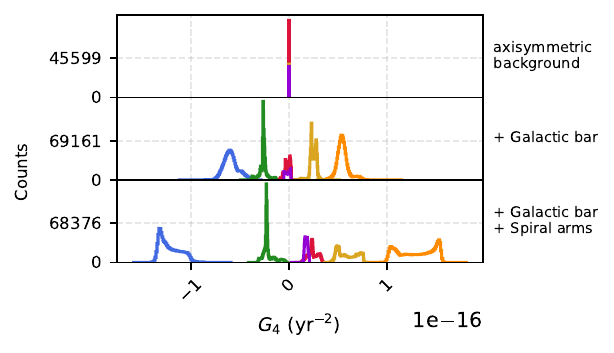}
    \includegraphics[width=1.0\textwidth]{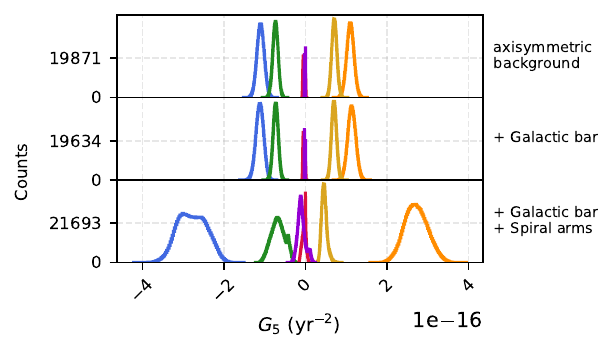}
    \includegraphics[width=1.0\textwidth]{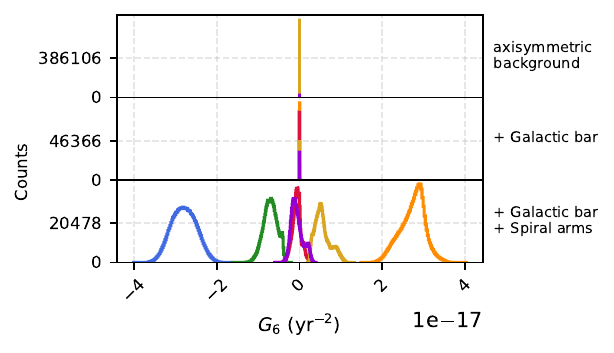}
    \caption{Continued.}
\end{figure}
\begin{figure}
    \includegraphics[width=1.0\textwidth]{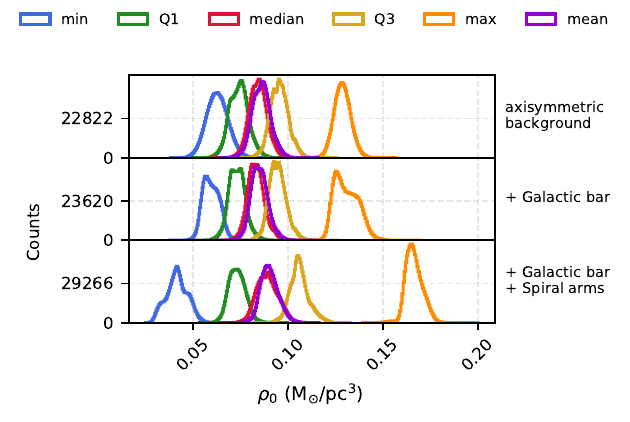}
    \caption{Same as Fig.~\ref{fig:Stats:histograms:R_z} but for trajectory-based summary statistics of the local Galactic mass density near the Sun.}
    \label{fig:Stats:histograms:rho}
\end{figure}
\begin{figure}
    \includegraphics[width=1.0\textwidth]{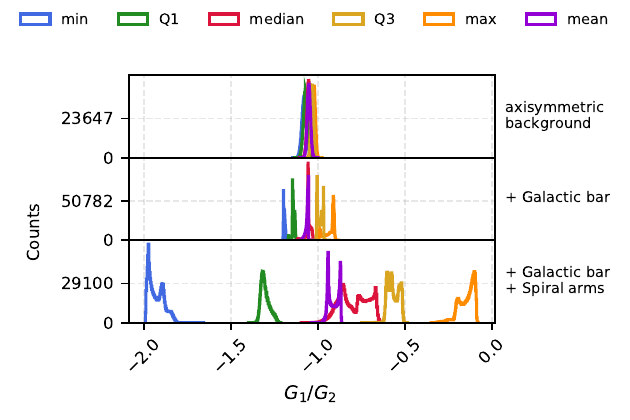}
    \includegraphics[width=1.0\textwidth]{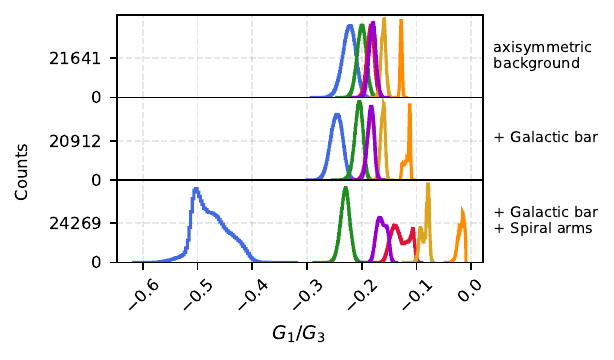}
    \includegraphics[width=1.0\textwidth]{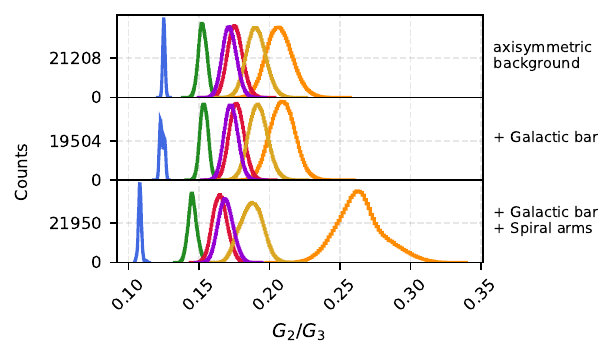}
    \caption{Same as Fig.~\ref{fig:Stats:histograms:R_z} but for trajectory-based summary statistics of the instantaneous ratios of the Galactic tide parameters.}
    \label{fig:Stats:histograms:ratios-G1_G2_G3}
\end{figure}
\addtocounter{figure}{-1}
\begin{figure}
    \includegraphics[width=1.0\textwidth]{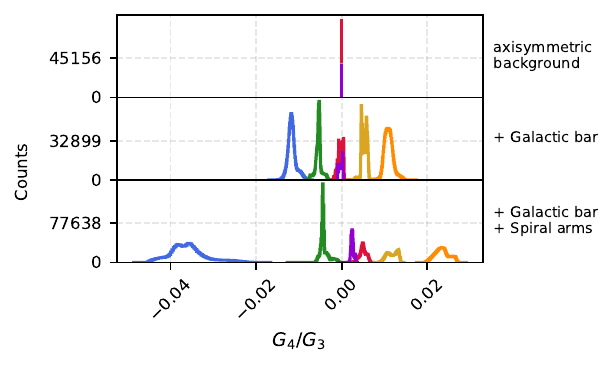}
    \includegraphics[width=1.0\textwidth]{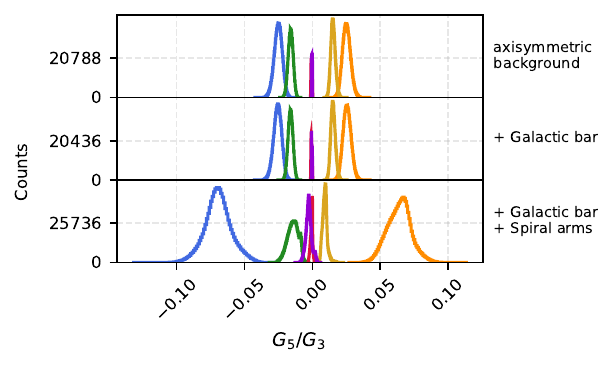}
    \includegraphics[width=1.0\textwidth]{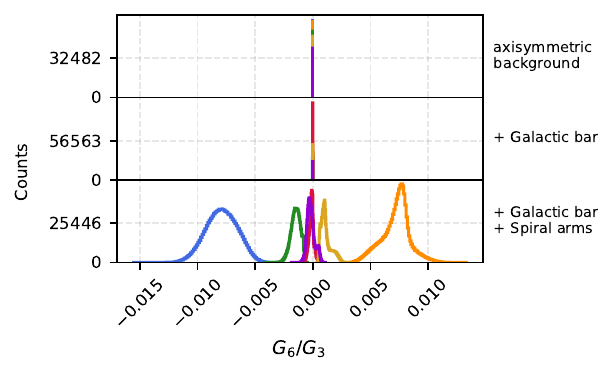}
    \caption{Continued.}
\end{figure}

In the three models, the first three quartiles of $R$ yielded by the integrations are nearly identical, meaning that the Sun spends roughly half of its dynamical evolution within the same range around the $R$-median (i.e. between Q1 and Q3). In the full model, however, the Sun spends less than half of its time (i.e., about $20\%$) exploring values of $R$ never reached in the axisymmetric model, with the minimum and maximum shifted by approximately $-0.5$~kpc and $+0.5$~kpc, respectively. Nevertheless, in all three models the Sun remains close to the Galactic plane and explores the same range of $z$, indicating that the non-axisymmetric component of the Galactic potential primarily affects the Sun's radial motion (see Fig.~\ref{fig:Stats:histograms:R_z}). From the geometry of Galactic tidal fields (see Fig.~\ref{fig:EGT:colormapG} in Sect.~\ref{sec:EGT2}) and the range of $R_\mathrm{s}$ (see Fig.~\ref{fig:Stats:histograms:R_z}), the Sun crosses the spiral arms many times. This explains why we observe a significant change in $G_1$, and to a lesser extent in $G_2$ and $G_3$, when moving from the axisymmetric model to the full model, because --- as noted in Sect.~\ref{sec:EGT2} --- only $G_1$ is strongly influenced by the spiral arms compared with $G_2$ and $G_3$. The changes observed in $G_2$ and $G_3$ are driven by the radial excursion of the Sun: $G_2$ increases as the Sun moves closer to the bar's region of influence, whereas $G_3$ follows the $\rho_0$ density distribution encountered by the Sun, which likewise increases toward the Galactic centre (see Figs.~\ref{fig:Stats:histograms:G1_G2_G3} and~\ref{fig:Stats:histograms:rho}). As expected, $G_4$ and $G_6$ vanish in the axisymmetric case, and the parameters $G_{4\text{---}5}$ are strongly affected by the addition of spiral arms. Finally, $G_6$ is also influenced by the spiral structure, which is the sole source of any variation in $G_6$ at the Sun's location; however, its range of variation is an order of magnitude smaller than that of $G_4$ and $G_5$ (see Fig.~\ref{fig:Stats:histograms:G1_G2_G3}).

Figure~\ref{fig:Stats:histograms:ratios-G1_G2_G3} follows the same method as Figs.~\ref{fig:Stats:histograms:R_z}, \ref{fig:Stats:histograms:G1_G2_G3}, and~\ref{fig:Stats:histograms:rho}, but for the following quantities: $G_1/G_2$, $G_1/G_3$, $G_2/G_3$, and $G_{4\text{---}6}/G_3$, which represent the instantaneous ratios of the Galactic tide parameters. Appendix~\ref{an:add-fig_hist} provides additional plots of other instantaneous ratios of the Galactic tide parameters. The histograms of $G_1/G_2$ are particularly informative, as they reveal the impact of potential axisymmetry breaking on the Galactic tides along the Sun's trajectory. For the full Galactic model, the ratio oscillates between approximately $0.1$ and $2$, corresponding to variations of $-90\%$ and $+100\%$, respectively, thus indicating a significant perturbation. This difference is reflected in the ratios $G_2/G_3$ and $G_1/G_3$, which can reach about~$0.26$ (median about~$0.16$) and about~$-0.5$ (median about~$-0.13$), respectively. On the other hand, the breaking is much less pronounced when the Galactic bar is the only non-axisymmetric component included. Consequently, in the region explored here by the Sun, the trends predicted by the axisymmetric case are the same as those obtained when only the Galactic bar is added. Finally, regardless of the model used, the approximations $|G_3|\gg |G_{1\text{---}2}|$ are only rough, since during most of the integration (and in some cases throughout the entire integration) we are in regions where $-G_1/G_3$ and $G_2/G_3$ exceed~$10\%$. In contrast, $G_3$ dominates $G_{4\text{---}6}$, whatever model is used.

\subsection{Discussion}\label{subsec:discussion}
The choice of initial conditions may be subject to discussion; however, in our case, we justify the choice made in Sect.~\ref{sec:Stats:orbits} for the reasons outlined therein. Nevertheless, we also performed the same numerical experiment with another set of initial velocity conditions:
\begin{equation}
    \begin{aligned}
       \mathbf{v}_{\mathrm{s},0}^{\text{alt.}}=\begin{pmatrix}-12.9^{\pm 3} & 250.2^{\pm 3} & 7.78^{\pm 0.6}\end{pmatrix}\,\text{km/s}\,.
    \end{aligned}
\end{equation}
The radial and vertical components are taken from~\cite{Drimmel-Poggio_2018}, except for the uncertainty on the vertical velocity, which we keep from~\cite{Schonrich-Binney-Dehnen_2010}, as we judge the uncertainty reported in~\cite{Drimmel-Poggio_2018} to be too small. The tangential component is unchanged, as it is consistent with our adopted value of the Sun’s initial radial position.

With this alternative set of initial conditions, we performed the same analysis as presented in Figs.~\ref{fig:EGT:signalex1} to~\ref{fig:Stats:histograms:ratios-G1_G2_G3}, as well as in the figures in Appendix~\ref{an:add-fig_hist}. The results obtained support the conclusions presented in this article. The figures related to this analysis are available in the `supplementary material' of this article.

\section{Summary and conclusion}\label{sec:cls}
Galactic tides are a fundamental component in understanding the dynamics of the most distant objects in the Solar System. They regulate the flux of observable long-period comets and produce oscillations of distant scattered disc objects. Moreover, they also play a role in related fields, such as the study of exoplanetary orbits~\citep{Veras-Evans_2013}; see Sect.~\ref{sec:intro} for another example. However, the current assumptions used in modelling Galactic tides --- namely, a circular stellar orbit in the Galactic plane, and an axisymmetric Galactic potential that is independent of time~\citep[see][]{Heisler-Tremaine_1986} --- are severe limitations and contrast with our current understanding of the Milky Way and the quality of observations obtained in recent years. As mentionned by~\citet[][Conclusions]{Veras-Evans_2013}, stars lead much more complex lives: They are usually inclined to the Galactic plane, their orbits are often eccentric and sometimes chaotic, and they experience perturbations that can move them by many kiloparsecs. It is in this context that we were led to develop a general expression for Galactic tides felt by a test particle in orbit around a star, for any trajectory of this star in any Galactic potential. We then applied it to the case of the Solar System, embedded in one of the most advanced Galactic models, which has been adjusted using Gaia DR3 data. The Galactic model is based on~\cite{Khalil_2025}, which we have extended to 3D in order to explore vertical oscillations in the Sun's orbit. In this way, we explored the possible backward motion of the Sun over the last gigayear by performing $10^{6}$~integrations with different initial conditions, studied the properties of the Galactic tide parameters, and then performed a statistical analysis of the general trends in the time evolution of Galactic tides along the Sun's orbit through the Galaxy. In order to determine the origin of the observed features, we compared the results obtained with three different configurations: (i)~using the complete Galaxy model; (ii)~using only its axisymmetric component; and (iii)~using the model without the spiral arms, retaining only the axisymmetric component and the Galactic bar.

A few caveats are worth mentioning. First, although the Galactic tide parameters have been estimated using the most recent model for the bar and spiral arms adjusted to Gaia data, this model may carry large systematic uncertainties. The spiral arm fit could correspond to a local minimum and, in any case, cannot necessarily be trusted over 1~Gyr. On longer timescales, recurrent cycles of spiral arms are anyway expected, and could have a substantial influence on the past motion of the Sun. This neglected effect of recurrent spirals largely overshadows another caveat, related to the influence of past stellar encounters on the solar orbit. Even at the axisymmetric level, the~\citet{Khalil_2025} model had been fitted to in-plane motions, but its systematic uncertainties in the vertical structure of the potential have not been explored or adjusted to data. However, our model predicts local surface mass densities in the solar neighbourhood that are in good agreement with the estimates of~\cite{Horta-PriceWhelan-Hogg-Johnston-Widrow-Dalcanton-Ness-Hunt_2024}. In particular, we find that the local surface density within a height of $1.1$~kpc is: $47.76\,{\rm M}_\odot/\text{pc}^2$ for baryons without spiral arms; $46.52\,{\rm M}_\odot/\text{pc}^2$ for baryons with spiral arms; $28.33\,{\rm M}_\odot/\text{pc}^2$ for the dark matter halo; $76.09\,{\rm M}_\odot/\text{pc}^2$ for baryons plus the dark matter halo without spiral arms; and $74.85\,{\rm M}_\odot/\text{pc}^2$ for baryons plus the dark matter halo with spiral arms. These values are consistent within the uncertainties reported in~\cite{Horta-PriceWhelan-Hogg-Johnston-Widrow-Dalcanton-Ness-Hunt_2024}. Nevertheless, neglecting the thick disc as we did in our model could slightly affect the vertical period of the Sun, even for a small $z_{\rm max}$, and all the more so if the model is applied to exoplanetary systems with larger vertical excursions and/or periods. Indeed, for stars located close to the Sun, the inclusion of a thick disc component in a Galactic model could increase the vertical excursions of an orbit by up to 40-50\%~\citep[see e.g.][]{Pouliasis-DiMatteo-Haywood_2017}. In that sense, the present paper should be considered as a first quantitative exploration of the variation of the Galactic tide parameters in a realistic non-axisymmetric gravitational potential of the Galaxy, but not as a computation of definitive values for these parameters. Nevertheless, let us note that the Galactic tide parameters less affected by the spiral arms, namely $G_2$ and $G_3$, should be considered more secure than $G_1$, $G_4$, $G_5$, and $G_6$, where the influence of the spiral arms is more pronounced.

The values of the Galactic tide parameters and their statistics outlined above can now be used to determine the validity of common assumptions made in previous works. For instance, they can be applied to study the impact of Galactic tides on Oort cloud comets and distant TNOs~\cite[see e.g.][]{Saillenfest-Fouchard-Ito-Higuchi_2019}, the dynamics of wide stellar triples in the Galaxy~\citep[][]{Grishin_Perets-2022}, or other related fields using Galactic tide models (see Sect.~\ref{sec:intro}). We find that the generalisation of the Galactic tides provides three new Galactic tide parameters, which we denote $G_4$, $G_5$, and $G_6$, and gives general expressions of the existing parameters $G_1$, $G_2$, and $G_3$. These expressions of $G_{1}$ to $G_6$ show that the appropriate form of the Galactic tide parameters depends on the application considered, since Galactic tides are intrinsic to both the stellar trajectory and the Galactic model. For instance, we may wonder how the conclusions of~\cite{Gardner-Nurm-Flynn-Mikkola_2011} on the flux of long-period comets would have been changed if they had taken into account the parameter $G_5$ (which actually takes significantly non-zero values). Finally, these general expressions for Galactic tide parameters enable the quantification of Galactic tides in complex scenarios of stellar motion, including stars in the Milky Way with orbits similar to, or highly atypical compared to, the Sun, as well as in the context of the Galaxy's dynamical evolution.

The results presented in this article indicate that, regardless of which Galactic model is used, for most applications, it appears questionable to assume constant values for the Galactic tides or a constant value for the local Galactic mass density in the neighbourhood of the Sun. In fact, the local Galactic mass density varies from roughly $0.05$ to $0.15$~$\mathrm{M}_\odot/\text{pc}^3$ for the axisymmetric case, and from roughly $0.03$ to $0.20$~$\mathrm{M}_\odot/\text{pc}^3$ for the full model, which represents a significant variation in both cases. These variations are in sharp contrast with the constant values used in most previous studies. For instance, models using a constant value of $\rho_0=0.10~\mathrm{M}_\odot/\mathrm{pc}^3$ to study the dynamics of Oort cloud comets~\citep[see e.g.][]{Levison-Dones-Duncan_2001} adopt a density that is somewhat higher than the median value predicted by our various galactic models. Moreover, temporary excursions to higher or lower values of~$\rho_0$ may noticeably affect the flux of long-period comets. The excursions might also affect the limits of the `inert zone' between the scattered disc and the Oort cloud and produce sporadic changes in the orbital elements of distant TNOs. In addition, the variation range of the Galactic tides in the full model is considerable: up to nearly one order of magnitude variation in $G_1$, $G_2$, and $G_3$. We have also shown that the common approximation made in previous studies~\citep[see e.g.][]{Matese-Whitmire_1996,Levison-Dones-Duncan_2001}, namely neglecting $G_1$ and $G_2$ relative to $G_3$ on the basis that they differ by an order of magnitude, is no longer valid, even in the axisymmetric case. Such departures from the $G_3$-dominated regime may break the quasi-integrable character usually assumed for the long-term tidal evolution of Oort-cloud comets, especially in the inner Oort cloud (i.e., semi-major axis $a \lesssim 20\,000\,\mathrm{au}$), and can therefore have significant consequences for the flux of long-period comets injected into the inner Solar System. The parameter $G_3$ is mainly affected by the fact that the range of galactocentric distance is wider for the non-axisymmetric models. Since the variation in the perihelion distance of comets over one orbital period under a vertical Galactic tide is proportional to $G_3 a^{7/2}$~\citep[see, e.g.][]{Fouchard-Froeschle-Rickman-Valsecchi_2010}, the variation of $G_3$ will directly affect the minimal semi-major axis for which a long period comet can be observable considering a vertical tide only. For instance, by considering the extremum values of $G_3$ obtained here, the minimum semi-major axis required for a comet to jump the Jupiter-Saturn barrier~\citep{Fouchard-Higuchi-Ito-Maquet_2018} varies roughly from~$17\,000$ to~$27\,000$~au. Finally, the introduction of the new Galactic parameters $G_4$, $G_5$, and $G_6$ may complicate the overall dynamics of long-period comets and the structure of the Oort cloud. We leave a full study of the effects of the non-$G_{3}$ tidal parameters on the long-term dynamics of Oort cloud comets to future work. They may also have a strong impact on extrasolar bodies having a large out-of-the-plane component in the Galaxy.

\section*{Data availability}
We provide with this paper a text file containing the tabulated nominal time evolution of all parameters (the Sun's position, velocity, local Galactic mass density, and Galactic tide parameters) as a function of time, for use in Solar System studies. We also provide an alternative tabulated nominal time evolution of these parameters, corresponding to a different set of initial conditions than those used in Sect.~\ref{sec:Stats:orbits} (see Sect.~\ref{subsec:discussion} for more details). Finally, we also include in the `supplementary material' figures analogous to those presented in Figs.~\ref{fig:EGT:signalex1} to~\ref{fig:Stats:histograms:ratios-G1_G2_G3}, as well as those in Appendix~\ref{an:add-fig_hist}, obtained using different initial conditions consistent with the Sun’s alternative nominal orbit presented in Sect.~\ref{subsec:discussion}. The `supplementary material' is available here: \url{https://doi.org/10.5281/zenodo.20933813}.

\begin{acknowledgements}
    We gratefully acknowledge the remarks made by the referee on this paper. This work was supported by the Programme de Planétologie (PNP) of CNRS/INSU, co-funded by CNES. Y.~R.~K. acknowledges funding from the Interdisciplinary Thematic Institute IRMIA++, as part of the ITI 2021–2028 program of the University of Strasbourg, CNRS and Inserm, supported by IdEx Unistra (ANR-10-IDEX-0002), and by the SFRI-STRAT’US project (ANR-20-SFRI-0012) under the framework of the French Investments for the Future Program. Y.~R.~K. was also supported by the European Research Council (ERC) under the European Union’s Horizon 2020 research and innovation programme (grant agreement No.~101117455). 
\end{acknowledgements}

\bibliographystyle{aa}
\bibliography{SunMotion}
\begin{appendix}
\section{The isochrone symplectic integrator}\label{an:Perf_int}
The Hamiltonian reproducing the equations of motion of the Sun orbiting within the Milky Way can be chosen as
\begin{equation}
    \label{eq:Perf_int:H}
    \begin{aligned}
        \mathcal{H}(\mathbf{q},\mathbf{p},t,T)=\frac{1}{2}\mathbf{p}^2+\Phi(\mathbf{q},t)+T\,,
    \end{aligned}
\end{equation}
where $\Phi$ is the Galactic potential, $\mathbf{p}=\dot{\mathbf{q}}$ is the momentum conjugate to the position $\mathbf{q}$, and $T$ is the momentum conjugate to the time $t$, introduced in order to obtain an autonomous system (only required for checking the conservation of the Hamiltonian value). For the test presented below, we used the full Galactic model described in Sect.~\ref{sec:GM}. 

We split the Hamiltonian into two parts,
\begin{equation}
    \label{eq:Perf_int:split}
    \begin{aligned}
        \mathcal{H}=\mathcal{A}+\varepsilon\mathcal{B}\,,
    \end{aligned}
\end{equation}
in order to use the isochrone splitting~\citep{Bougakov-Saillenfest-Fouchard_2025},
\begin{align}
    \label{eq:A-HAM}
    \mathcal{A}(\mathbf{q},\mathbf{p},T) &= \frac{1}{2}\mathbf{p}^2 + \Psi(\mathbf{q})+T\,,\\
    \label{eq:eB-HAM}
    \varepsilon\mathcal{B}(\mathbf{q},t) &= \Phi(\mathbf{q},t) - \Psi(\mathbf{q})\,,
\end{align}
where
\begin{equation}
    \label{eq:Perf_int:isochrone}
    \begin{aligned}
        \Psi(\mathbf{q};\kappa,b)=\frac{-\kappa}{b+\sqrt{\mathbf{q}^2+b^2}}\,,
    \end{aligned}
\end{equation}
is the isochrone potential~\citep{Henon_1959_1}. This approach relies on isochrone drifts and velocity kicks\footnote{A simple example implementation is available at \url{https://github.com/AstroLexandre/IsochroneSymplecticIntegrator_QuickExample}, which includes the analytic isochrone propagator used to evolve the isochrone Hamiltonian flow~\citep{Bougakov-Saillenfest-Fouchard_2026}.}. The splitting depends on two constant parameters, $\kappa$ and $b$, which must be chosen such that the absolute value of the ratio between the split parts in Eq.~\eqref{eq:Perf_int:split}, $\varepsilon$, is as small as possible. When $b=0$ (and $\kappa\neq0$), it reduces to Kepler drifts and velocity kicks, corresponding to the Kepler splitting; in this case, the parameter $\kappa$ and the coordinates $\mathbf{q}$ are chosen to give the best performance (see, e.g.~\citealp{Farres-etal_2013,Hernandez_2017,Rein_2019_1}). Thus, the isochrone splitting can be seen as a generalisation of Kepler splitting. In the limiting case $\Psi=0$ (i.e. $\kappa=0$), it reduces to linear drifts and velocity kicks, corresponding to the kinetic splitting, as used, for instance, in the standard Leapfrog integrator~\citep{Ruth_1983}.

\begin{figure}
    \includegraphics[width=0.95\columnwidth]{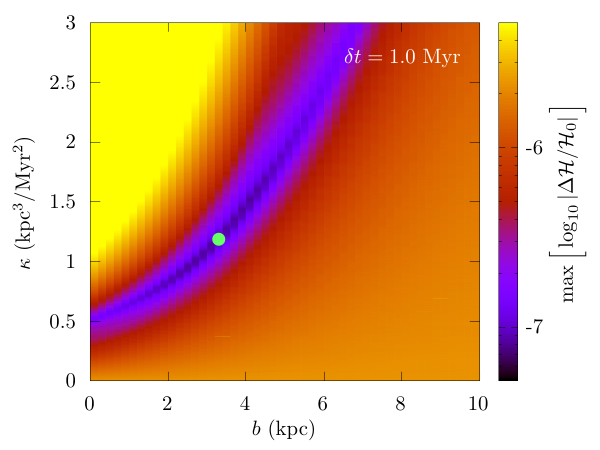}
    \caption{Optimising the parameterisation of isochrone splitting. The green dot corresponds to one of the optimal values for the isochrone splitting parameters.}
    \label{fig:Perf_int:kbmap}
\end{figure}

We denote by $\mathcal{H}_0$ the Hamiltonian at the initial time of the integration, and by $\Delta\mathcal{H}$ its instantaneous variation relative to $\mathcal{H}_0$ at time $t$; perfect conservation implies that $\Delta\mathcal{H}=0$ at all times during the integration. Figure~\ref{fig:Perf_int:kbmap} shows how to choose the values of the isochrone splitting parameters, $\kappa$ and $b$, in order to optimise the integration with the $\mathcal{SABA}_2$ symplectic integrator of~\cite{Laskar-Robutel_2001}; we denote the optimal values of the isochrone splitting parameters as $\kappa_\star$ and $b_\star$. The integrations were performed over $1$~Gyr with a time step of $\delta t=1$~Myr.

\begin{figure}
    \includegraphics[width=0.95\columnwidth]{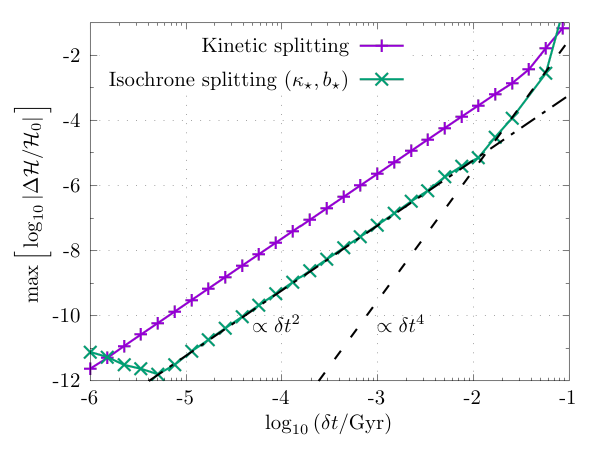}
    \caption{Maximum absolute relative Hamiltonian variation versus the integration time step (both splitting schemes are combined with $\mathcal{SABA}_2$).}
    \label{fig:Perf_int:timestep}
\end{figure}

\begin{figure}
    \includegraphics[width=0.95\columnwidth]{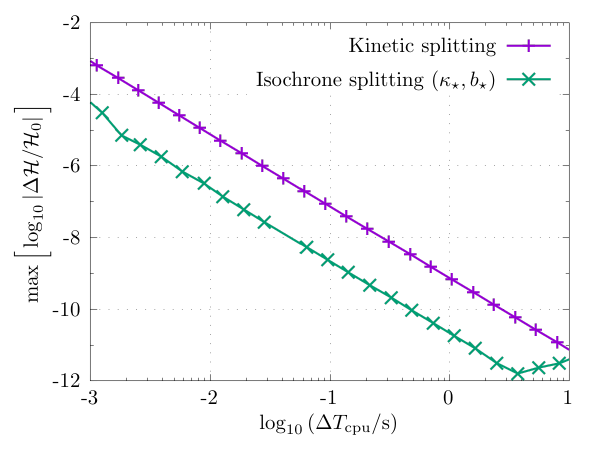}
    \caption{Maximum absolute relative Hamiltonian variation versus CPU integration time.}
    \label{fig:Perf_int:speed}
\end{figure}

The $\mathcal{SABA}_2$ integrator has errors of order $\mathcal{O}(\varepsilon\delta t^4)+\mathcal{O}(\varepsilon^2\delta t^2)$. This implies that the integrator is fourth-order in $\delta t$ in the best case and only second-order in the worst case, depending on the size of $\varepsilon$. For moderate constraint in energy conservation (relative error $10^{-3}$) or stronger, Figs.~\ref{fig:Perf_int:timestep} and~\ref{fig:Perf_int:speed} show that the isochrone integrator outperforms the standard Leapfrog-like integrator. In fact, for the same level of Hamiltonian conservation, the isochrone splitting with the parameterisation $\kappa=\kappa_\star$ and $b=b_\star$ allows larger time steps than the standard kinetic splitting. As a result, the integration is faster than with the standard kinetic splitting.


\section{Comparison between $G_1$ and $G_5$ in another axisymmetric potential}\label{an:add-fig_comp}
\begin{figure}
    \includegraphics[width=0.9\columnwidth]{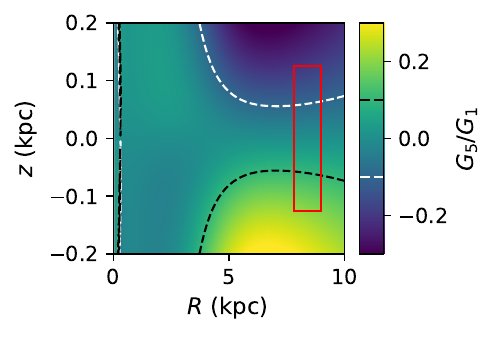}
    \caption{Ratio of $G_5/G_1$ in the axisymmetric potential of~\cite{Gardner-Nurm-Flynn-Mikkola_2011}. As in Fig.~\ref{fig:EGT:ratioG5onG1} in Sect.~\ref{sec:EGT1}, the white and black curves correspond to values of about $-10\%$ and $+10\%$, respectively. The red rectangle delineates the region explored by the Sun's orbit as derived in~\cite{Gardner-Nurm-Flynn-Mikkola_2011}.}
    \label{fig:EGT:ratioG5onG1_Gardner}
\end{figure}

Figure~\ref{fig:EGT:ratioG5onG1_Gardner} is analogous to Fig.~\ref{fig:EGT:ratioG5onG1} in Sect.~\ref{sec:EGT1}, but is based on the axisymmetric potential of~\cite{Gardner-Nurm-Flynn-Mikkola_2011}. We find that the Sun's orbit derived in their model explores regions of the $(R,z)$-plane where the ratio $|G_5/G_1|$ reaches values up to $20\%$.

\section{Supplementary plots of Galactic tidal fields}\label{an:add-fig_color}
\begin{figure*}
    \includegraphics[width=0.92\columnwidth]{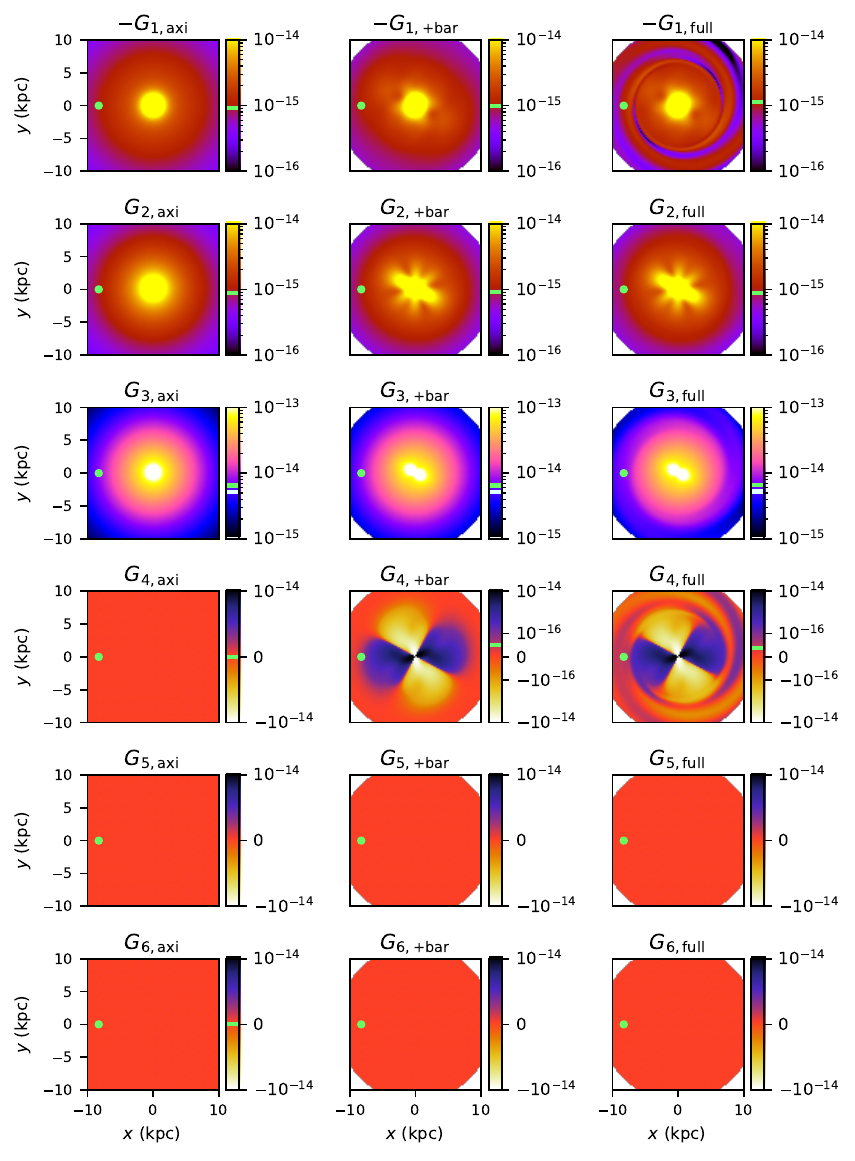}
    \caption{As in Fig.~\ref{fig:EGT:colormapG} in Sect.~\ref{sec:EGT2}, except that the Galactic tidal fields are evaluated in the Galactic plane ($z=0$).}
    \label{fig:add-fig_color:colormapG-1}
\end{figure*}

\begin{figure*}
    \includegraphics[width=0.92\columnwidth]{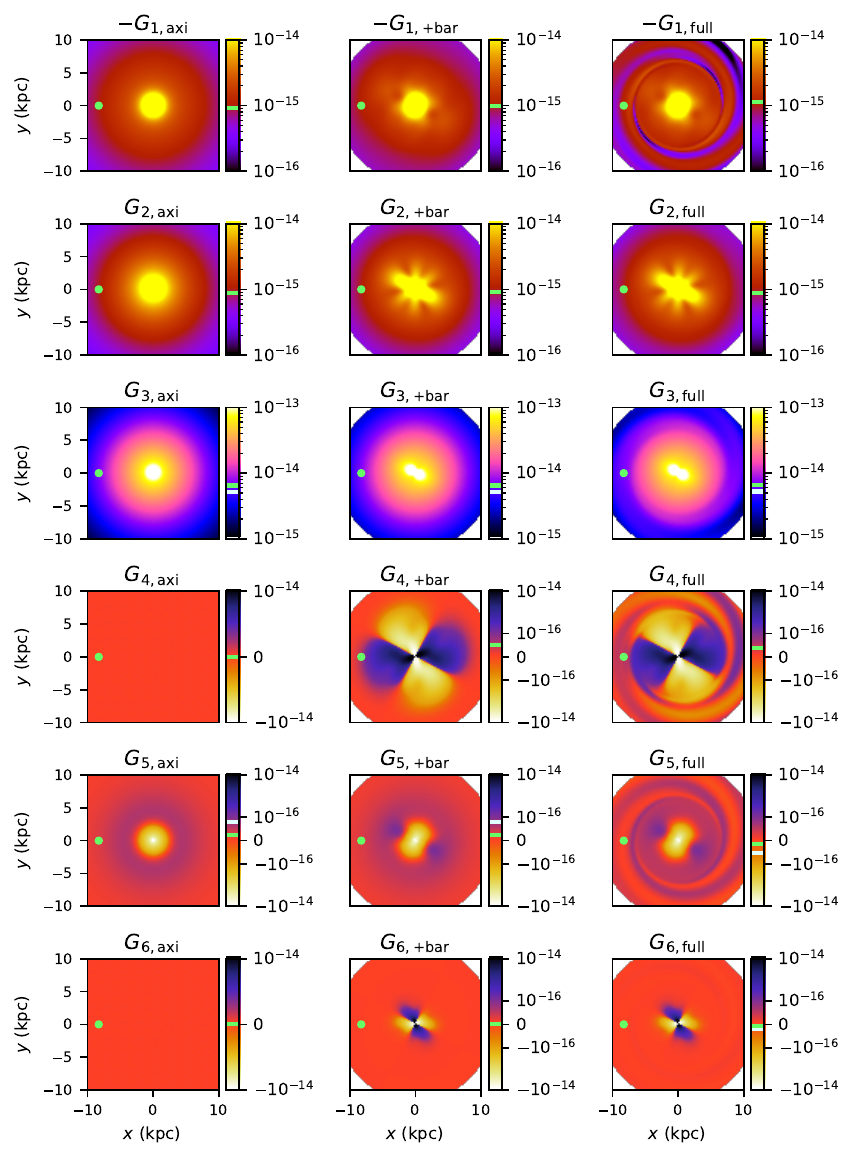}
    \caption{As in Fig.~\ref{fig:EGT:colormapG} in Sect.~\ref{sec:EGT2}, except that here $z=20$~pc, so that the green dot and the slices lie in the same plane.}
    \label{fig:add-fig_color:colormapG-2}
\end{figure*}

Figures~\ref{fig:add-fig_color:colormapG-1} and~\ref{fig:add-fig_color:colormapG-2} are analogous to Fig.~\ref{fig:EGT:colormapG} in Sect.~\ref{sec:EGT2}, except that they show snapshots computed in the Galactic plane (i.e. $z=0$) and in the same $z$-plane as the Sun (i.e. $z=20$~pc), respectively. These figures confirm the analysis made in Sect.~\ref{sec:EGT2}.


\section{Supplementary histogram plots}\label{an:add-fig_hist}
Figure~\ref{fig:add-fig:histograms:ratios} is a continuation of Fig.~\ref{fig:Stats:histograms:ratios-G1_G2_G3} in Sect.~\ref{sec:Stats:results}, which shows the statistics of the instantaneous ratios $G_{4\text{---}6}/G_{1\text{---}2}$. These figures show that the hierarchy between the non-$G_3$ diagonal terms (i.e. $G_1$ and $G_2$) and the non-$G_6$ off-diagonal terms (i.e. $G_4$ and $G_5$) disappears when spiral arms are taken into account in the Galactic model.

\begin{figure}
    \includegraphics[width=1.0\textwidth]{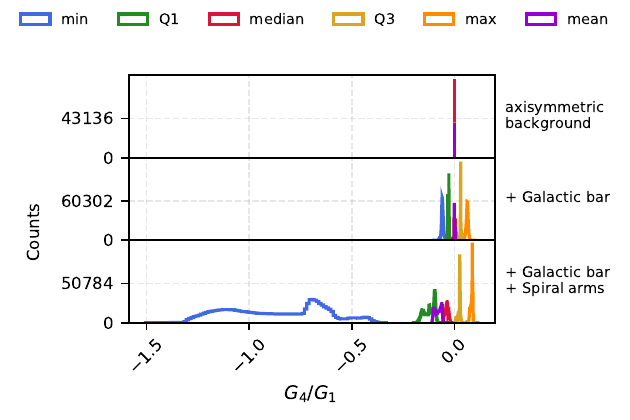}
    \includegraphics[width=1.0\textwidth]{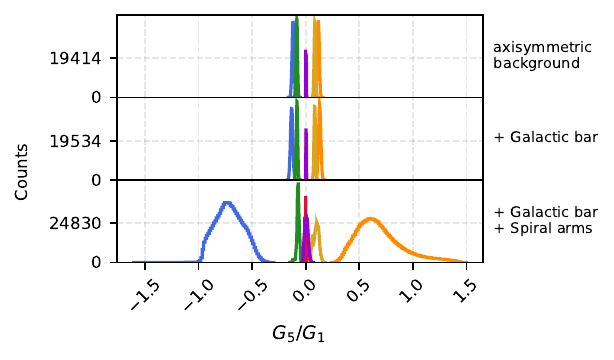}
    \includegraphics[width=1.0\textwidth]{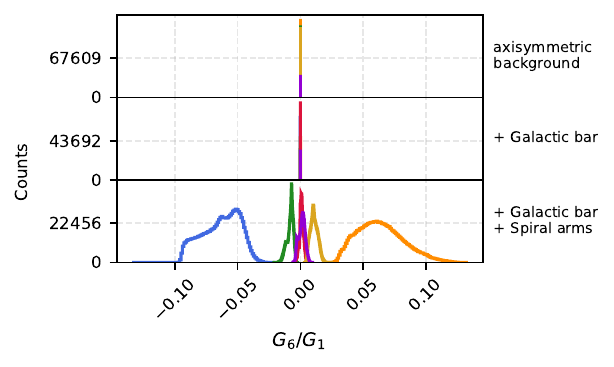}
    \caption{Same legend as in Fig.~\ref{fig:Stats:histograms:ratios-G1_G2_G3} in Sect.~\ref{sec:Stats:results}.}
    \label{fig:add-fig:histograms:ratios}
\end{figure}
\addtocounter{figure}{-1}
\begin{figure}
    \includegraphics[width=1.0\textwidth]{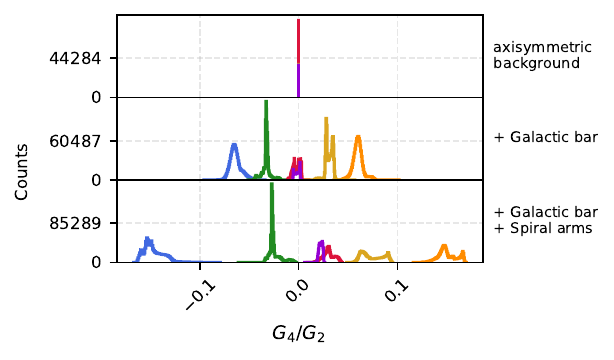}
    \includegraphics[width=1.0\textwidth]{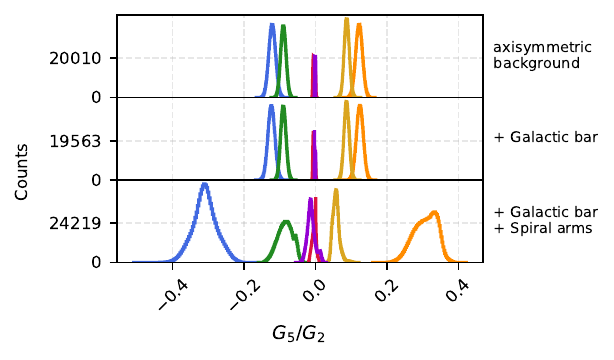}
    \includegraphics[width=1.0\textwidth]{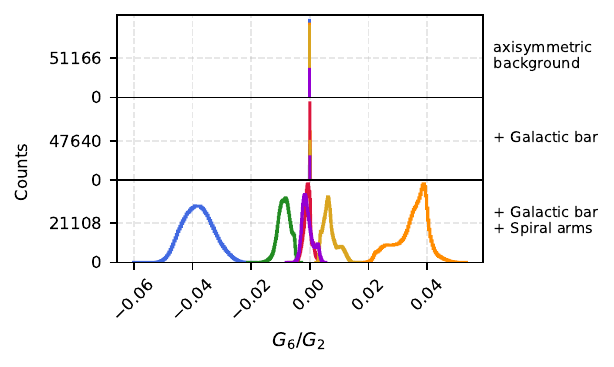}
    \caption{Continued.}
\end{figure}
\end{appendix}

\end{document}